\pdfoutput=1

\documentclass[%
reprint,
 superscriptaddress,
 amsmath,amssymb,
 aps,
]{revtex4-1}

\usepackage{graphicx}%
\usepackage{dcolumn}%
\usepackage{bm}%

\usepackage[normalem]{ulem}
\usepackage[svgnames,psnames]{xcolor} %
\usepackage[colorlinks,citecolor=DarkGreen,linkcolor=FireBrick,linktocpage,unicode]{hyperref} %
\usepackage{textcomp}  %
\usepackage{rotating}  %

\newcommand{\relmiddle}[1]{\mathrel{}\middle#1\mathrel{}}
\newcommand{\set}[2]{\left\{ #1 \relmiddle| #2 \right\}}
\DeclareMathOperator{\morethan}{>}

\begin{document}

\preprint{APS/123-QED}

\title{Finding well-optimized special quasirandom structures with decision diagram}

\author{Kohei Shinohara}
    \email{shinohara@cms.mtl.kyoto-u.ac.jp}
    \affiliation{Department of Materials Science and Engineering, Kyoto University, Kyoto 606-8501, Japan}
\author{Atsuto Seko}
    \email{seko@cms.mtl.kyoto-u.ac.jp}
    \affiliation{Department of Materials Science and Engineering, Kyoto University, Kyoto 606-8501, Japan}
    \affiliation{Center for Elements Strategy Initiative for Structure Materials (ESISM), Kyoto University, Kyoto 606-8501, Japan}
\author{Takashi Horiyama}
    \affiliation{Faculty of Information Science and Technology, Hokkaido University, Sapporo 060-0814, Japan}
\author{Isao Tanaka}
    \affiliation{Department of Materials Science and Engineering, Kyoto University, Kyoto 606-8501, Japan}
    \affiliation{Center for Elements Strategy Initiative for Structure Materials (ESISM), Kyoto University, Kyoto 606-8501, Japan}
    \affiliation{Nanostructures Research Laboratory, Japan Fine Ceramics Center, Nagoya 456-8587, Japan}

\date{\today}%

\begin{abstract}
The advanced data structure of the zero-suppressed binary decision diagram (ZDD) enables us to efficiently enumerate nonequivalent substitutional structures.
Not only can the ZDD store a vast number of structures in a compressed manner, but also can a set of structures satisfying given constraints be extracted from the ZDD efficiently.
Here, we present a ZDD-based efficient algorithm for exhaustively searching for special quasirandom structures (SQSs) that mimic the perfectly random substitutional structure.
We demonstrate that the current approach can extract only a tiny number of SQSs from a ZDD composed of many substitutional structures ($\morethan 10^{12}$).
As a result, we find SQSs that are optimized better than those proposed in the literature.
A series of SQSs should be helpful for estimating the properties of substitutional solid solutions.
Furthermore, the present ZDD-based algorithm should be useful for applying the ZDD to the other structure enumeration problems.
\end{abstract}

\maketitle

\section{Introduction}
Modeling a perfectly random structure using periodic structures has been performed for several decades to estimate the physical properties of substitutional solid solutions using density functional theory (DFT) calculations.
One approach adopts special quasirandom structures (SQSs) \cite{SQS1}, which are periodic structures almost identical to a perfectly random structure in terms of correlation functions.
SQSs have been widely used with DFT calculations to estimate physical properties of perfectly random structures such as formation enthalpy \cite{WOLVERTON20013129,PhysRevB.75.045123,PhysRevB.91.024106,doi:10.1021/jacs.6b03207}, lattice distortion \cite{PhysRevApplied.2.044009,PhysRevMaterials.1.023404}, elastic properties \cite{PhysRevB.81.094203}, electronic properties including the band gap \cite{PhysRevLett.76.664,PhysRevB.84.081102,Hinuma2016}, paramagnetic properties \cite{PhysRevLett.92.185702,PhysRevB.85.125104,PhysRevB.90.134106,doi:10.1063/1.4918996}, and piezoelectric properties \cite{PhysRevLett.104.137601}.

In a rigorous way, the periodic structures that mimic a perfectly random structure are exhaustively explored from a vast number of nonequivalent substitutional structures called derivative structures \cite{Buerger1947}.
However, the size of the entire set of derivative structures increases exponentially with the number of representative atoms included in derivative structures and the number of atom types.
A stochastic approach that can approximate SQSs with moderate precision, such as simulated annealing, has also been employed \cite{mcsqs,PhysRevB.76.144204,JIANG20094716}.
The stochastic approach usually generates a reasonable SQS close to the perfectly random structure for a given SQS size and given set of clusters.
On the other hand, the stochastic method does not necessarily generate the best SQSs.
Recently, a small set of ordered structures (SSOSs) \cite{SSOS,PhysRevB.102.174209}, which replaces a perfectly random structure with a weighted average of several periodic structures, has also been proposed.
These approaches as well as the exhaustive search of SQSs should be helpful.

Recently, a compact data structure of the zero-suppressed binary decision diagram (ZDD) \cite{Minato1993} has been applied to the efficient enumeration of derivative structures \cite{doi:10.1063/5.0021663}.
The ZDD enables us to store many structures in a compressed manner and enumerate more than $10^{15}$ derivative structures in a reasonable time.
Another advantage of using the ZDD is that a set of structures satisfying given constraints can be extracted from the ZDD efficiently, and the extraction of only nonequivalent structures is such a constraint \cite{doi:10.1063/5.0021663}.
In the field of discrete algorithms, the ZDD has been used to enumerate subgraphs of a given graph with a specific property, such as all self-avoiding paths between two vertices (s-t paths) \cite{knuth2009art,KAWAHARA2017}.
This feature, which efficiently narrows the list of derivative structures, is crucial in the practical use of enumerated derivative structures.

The extraction of SQSs from the ZDD is also regarded as a constraint for the enumeration of derivative structures.
In this study, we present an efficient algorithm for extracting SQSs from the ZDD composed of many substitutional structures without explicitly enumerating all derivative structures.
Using the ZDD-based algorithm, we find SQSs that are optimized better than those proposed in the literature, which should be helpful for estimating the properties of substitutional solid solutions.
This study should also help establish other ZDD-based algorithms extracting a limited number of feasible structures from many structures, such as a small number of candidates for ground-state and metastable structures.

This paper is organized as follows:
Section~\ref{sec:terminology} introduces the terminology for representing substitutional structures.
Section~\ref{sec:formulation} formulates a combinatorial problem for searching for SQSs by constructing a set of derivative structures satisfying several constraints.
Section~\ref{sec:algorithm} presents an algorithm for solving the problem with the ZDD.
Section~\ref{sec:results} shows the application of the present ZDD-based method for SQSs for face-centered cubic (fcc) and hexagonal close-packed (hcp) structures in binary, ternary, and quaternary systems.

\section{\label{sec:terminology}Terminology}

\begin{figure}[tb]
    \centering
    \includegraphics[width=\linewidth]{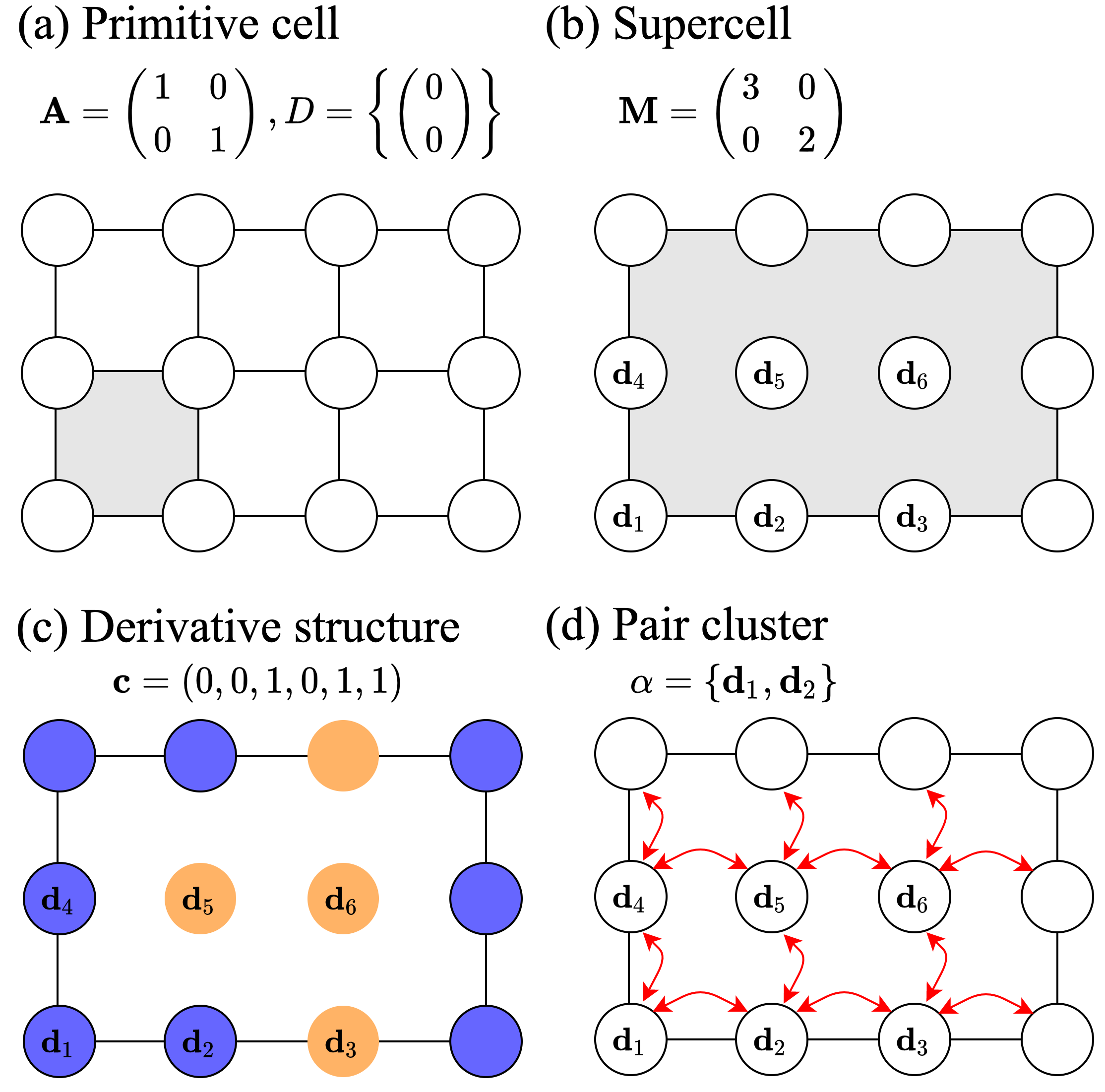}
    \caption{
        Illustration of two-dimensional derivative structure and pair clusters.
        (a) Primitive cell with basis vectors $\mathbf{A}$ and a set of point coordinates in the primitive cell, $D$.
        The shaded area represents the primitive cell.
        (b) Supercell derived from the primitive cell in (a).
        The index of its transformation matrix $\mathbf{M}$ is six, and there are six sites in the supercell, $D_{\mathbf{M}} = \{ \mathbf{d}_{1}, \mathbf{d}_{2}, \mathbf{d}_{3}, \mathbf{d}_{4}, \mathbf{d}_{5}, \mathbf{d}_{6} \}$.
        (c) Binary derivative structure with the supercell in (b).
        When integers 0 and 1 represent the blue and yellow atoms, respectively, the reduced labeling $\mathbf{c} = (0, 0, 1, 0, 1, 1)$ indicates that the blue atoms occupy sites $\mathbf{d}_{1}$, $\mathbf{d}_{2}$, and $\mathbf{d}_{4}$, and that the yellow ones occupy the other sites $\mathbf{d}_{3}$, $\mathbf{d}_{5}$, and $\mathbf{d}_{6}$.
        (d) Pair cluster $\alpha = \{ \mathbf{d}_{1}, \mathbf{d}_{2} \}$ corresponding to the nearest-neighbor pair and its symmetrically equivalent pair clusters $[\alpha]$, represented by the red arrows.
        In the supercell in (b), there are $|[\alpha]| = 12$ pair clusters equivalent to $\alpha$ due to the symmetry of the primitive cell.
    }
    \label{fig:terminology}
\end{figure}

Before we show the details of our algorithm for searching for SQSs, we briefly define the terminology to represent substitutional structures.
Figure~\ref{fig:terminology} illustrates the terminology defined in this section with a two-dimensional example.
It is straightforward to generalize it to arbitrary dimensions.

A primitive cell in three dimensions is specified with its basis vectors $\mathbf{A} = (\mathbf{a}_{1}, \mathbf{a}_{2}, \mathbf{a}_{3} )$ and point coordinates $D$, as shown in Fig.~\ref{fig:terminology}~(a).
We consider a $k$-ary substitutional structure derived from the primitive cell and label the atoms in the supercell with $k$ integers $\{ 1, \dots, k \}$.
A set of basis vectors of the supercell is written as $\mathbf{AM}$, where we refer to $\mathbf{M}$ as a transformation matrix.
We call the determinant of $\mathbf{M}$ the index of the supercell.
There are $|D| \cdot \det \mathbf{M}$ point coordinates in the supercell and we denote them as $D_{\mathbf{M}} = \{ \mathbf{d}_{1}, \dots, \mathbf{d}_{|D_{\mathbf{M}}|} \}$.
Figure~\ref{fig:terminology}~(b) shows a transformation matrix and the corresponding supercell in the two-dimensional example.
In the following, we proceed to a discussion with a fixed transformation matrix $\mathbf{M}$ and do not explicitly show the dependence of $\mathbf{M}$ in our notation as long as there is no ambiguity.

A $k$-ary derivative structure \cite{Buerger1947} is defined as a symmetrically nonequivalent substitutional structure.
The $k$-ary derivative structure can be specified with a set of occupation numbers and a transformation matrix of a supercell, $\mathbf{M}$.
We write the occupation number of atom type $p$ at site $i$ as
\begin{align}
    \tilde{c}_{i, p} =
    \begin{cases}
      1 & (\mbox{site $i$ is occupied with atom type $p$}) \\
      0 & (\mbox{otherwise})
    \end{cases}.
\end{align}
The sum of occupation numbers at each site $i$ over atom types should be one,
\begin{align}
    \label{eq:one-hot}
    \sum_{p=1}^{k} \tilde{c}_{i, p} = 1 \quad (i = 1,\dots,|D_{\mathbf{M}}|).
\end{align}
We refer to the set of occupation numbers as labeling and express it as
\begin{align}
    \label{eq:labeling}
    \tilde{\mathbf{c}} = \left( \tilde{c}_{1, 1}, \dots, \tilde{c}_{1, k}, \dots, \tilde{c}_{|D_{\mathbf{M}}|, 1}, \dots, \tilde{c}_{|D_{\mathbf{M}}|, k} \right).
\end{align}

We define the concentration of labeling $\tilde{\mathbf{c}}$ of atom type $p$ for site $I$ in the primitive cell as the average of occupation numbers over all sites translationally equivalent to site $I$, $[I]$.
This is written as
\begin{align}
    x_{I, p}(\tilde{\mathbf{c}}) = \frac{1}{|\det \mathbf{M}|} \sum_{j \in [I] } \tilde{c}_{j, p}.
\end{align}
Note that sites in the primitive cell are represented with capital symbols $I,\, J,\, \dots$ to distinguish them from sites in the supercell.
The sum of concentrations taken over atom types for each site $I$ should be one,
\begin{align}
    \sum_{p=1}^{k} x_{I, p}(\tilde{\mathbf{c}}) &= 1 \quad (I = 1, \dots, |D|).
\end{align}

A set of point coordinates is called a cluster.
We focus on pair cluster $\alpha$ and its correlation function.
We denote a set of symmetrically equivalent pair clusters with $\alpha$ as $[\alpha] \, (\in D_{\mathbf{M}} \times D_{\mathbf{M}})$.
Figure~\ref{fig:terminology}~(d) shows the first nearest-neighbor (NN) pair and its symmetrically equivalent pairs in the two-dimensional example.
We define the correlation function of pair cluster $\alpha$ between atom types $p$ and $q$ as
\begin{align}
    \label{eq:correlation_function}
    &\Pi_{ \alpha }^{pq} (\tilde{\mathbf{c}}) \nonumber \\
    &=
    \begin{cases}
        \frac{1}{ |[\alpha]| } \sum_{ (\mathbf{d}_{i}, \mathbf{d}_{j} ) \in [\alpha] } \tilde{c}_{i, p} \tilde{c}_{j, p} & (p = q) \\
        \frac{1}{ |[\alpha]| } \sum_{ (\mathbf{d}_{i}, \mathbf{d}_{j} ) \in [\alpha] } \left( \tilde{c}_{i, p} \tilde{c}_{j, q} + \tilde{c}_{i, q} \tilde{c}_{j, p} \right) & (p \neq q) \\
    \end{cases}
    .
\end{align}

In a binary system $(k = 2)$, only the occupation numbers of one atom types are needed to represent a derivative structure.
We denote the reduced labeling as
\begin{align}
    \mathbf{c}
        = (c_{1}, \dots, c_{|D_{\mathbf{M}}|})
        = (\tilde{c}_{1, 2}, \tilde{c}_{2, 2}, \dots, \tilde{c}_{|D_{\mathbf{M}}|, 2}).
\end{align}
Figure~\ref{fig:terminology}~(c) shows a binary derivative structure with the labeling representation $\mathbf{c}$.
We define the concentration of reduced labeling $\mathbf{c}$ as the average of occupation numbers over all sites equivalent to site $I$,
\begin{align}
    x_{I}(\mathbf{c}) = \frac{1}{|\det \mathbf{M}|} \sum_{j \in [I] } c_{j}.
\end{align}
The correlation function for reduced labeling $\mathbf{c}$ can be defined similarly to Eq.~\eqref{eq:correlation_function}.
For example, the correlation function of the pair cluster shown in Fig.~\ref{fig:terminology}~(d) is written as
\begin{align}
    \Pi_{ \alpha }(\mathbf{c})
    &= \frac{1}{12} (
        c_{1} c_{2} + c_{1} c_{3} + 2 c_{1} c_{4} + c_{2} c_{3}
        \nonumber \\
        &+ 2 c_{2} c_{5} + 2 c_{3} c_{6}
        + c_{4} c_{5} + c_{4} c_{6} + c_{5} c_{6}
    ).
\end{align}

Note that correlation functions of clusters are often formulated with pseudo-spin variables $\sigma_{i} = \pm 1$ but not occupation numbers.
Because each of these two variables can be converted to the other with a Vandermonde matrix \cite{FONTAINE199433}, there is no essential difference in which variables to use.
In this paper, we choose the definition of correlation functions represented with the occupation numbers because it is suitable for being regarded as a Boolean value of a combinatorial problem.

\section{\label{sec:formulation}Formulation of Searching for SQS}

\begin{figure}[tb]
    \centering
    \includegraphics[width=\linewidth]{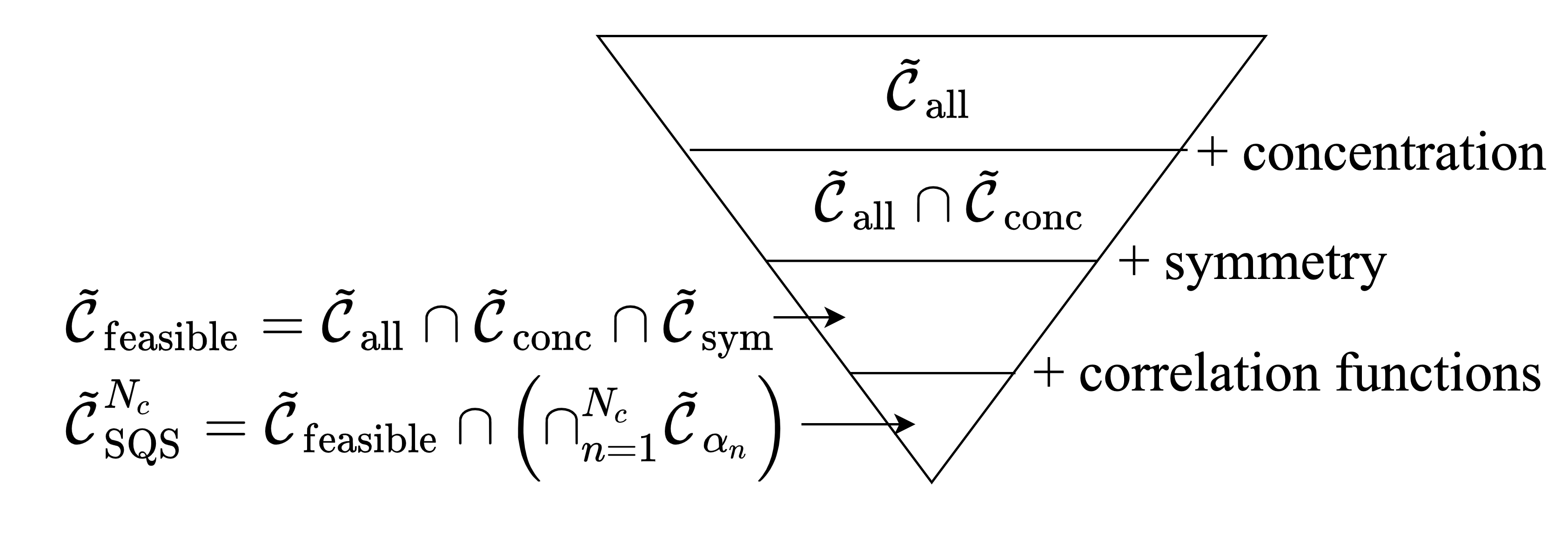}
    \caption{
        Schematic diagram of the searching policy of SQSs in this paper.
        The definitions of these sets are given in Sec.~\ref{sec:formulation}.
    }
    \label{fig:schematic}
\end{figure}

We formulate the combinatorial problem to find SQSs as enumerating labelings $\tilde{\mathbf{c}}$ satisfying Boolean constraints.
For a given supercell with transformation matrix $\mathbf{M}$, we consider the set of all possible labelings and eliminate labelings by imposing constraints that SQSs should satisfy.
Figure~\ref{fig:schematic} shows a schematic diagram of the labeling elimination.
Here, we introduce a constraint for eliminating labelings different from the perfectly random structure in terms of the concentration, a constraint for eliminating equivalent labelings in terms of the symmetry of the supercell, and a constraint for eliminating labelings different from the perfectly random structure in terms of correlation functions.
We define the set of all possible labelings $\tilde{\mathcal{C}}_{\mathrm{all}}$ and the set of possible labelings with the same concentration as the perfectly random structure, $\tilde{\mathcal{C}}_{\mathrm{conc}}$, in Sec.~\ref{sec:formulation-conc}.
In Sec.~\ref{sec:formulation-feasible}, we then define the set of feasible labelings $\tilde{\mathcal{C}}_{\mathrm{feasible}}$, which are nonequivalent labelings out of $\tilde{\mathcal{C}}_{\mathrm{conc}}$.
In Sec.~\ref{sec:formulation-corr}, we finally show the set of labelings composed of SQSs, $\tilde{\mathcal{C}}_{\mathrm{SQS}}^{N_{c}}$, which are the same as the perfectly random structure in terms of correlation functions.

\subsection{\label{sec:formulation-conc}Concentration}

Since the sum of occupation numbers at each site over atom types is one [Eq.~\eqref{eq:one-hot}], all possible labelings in a $k$-ary system should belong to the following set:
\begin{align}
    \label{eq:set-one-hot}
    \tilde{\mathcal{C}}_{\mathrm{all}}
        = \set{ \tilde{\mathbf{c}} \in \{ 0, 1 \}^{ k |D_{\mathbf{M}}|} }{ \sum_{p=1}^{k} \tilde{c}_{i, p} = 1 \quad (\forall i) }.
\end{align}
When we consider a $k$-ary perfectly random structure whose concentration of atom type $p$ at site $I$ is $\overline{x}_{I, p}$, a substitutional structure that mimics the perfectly random structure is desired to have the same concentration as $\overline{x}_{I, p}$.
For simplicity, we assume the concentrations at symmetrically equivalent sites in the primitive cell are equal.
Thus, its labeling representation $\tilde{\mathbf{c}}$ belongs to the following set
\footnote{
    The constraint in Eq.~\eqref{eq:set_concentration} is stricter than the composition constraint in Ref.~\onlinecite{doi:10.1063/5.0021663}.
    The former fixes the concentration of each site, whereas the latter only fixes the averaged concentration over all sites in a supercell.
    The former constraint should be more appropriate to search for the substitutional structure closest to the perfectly random structure.
}:
\begin{align}
    \label{eq:set_concentration}
    \tilde{\mathcal{C}}_{\mathrm{conc}}
    = \set{ \tilde{\mathbf{c}} \in \{ 0, 1 \}^{ k |D_{\mathbf{M}}|} }{ x_{I, p}(\tilde{\mathbf{c}}) = \overline{x}_{I, p} \quad (\forall I, p) }.
\end{align}
When we consider a binary perfectly random structure with concentration $\overline{x}_{I}$, the reduced labeling $\mathbf{c}$ corresponding to a substitutional structure that mimics the perfectly random structure belongs to the following set
\begin{align}
    \mathcal{C}_{\mathrm{conc}} = \set{ \mathbf{c} \in \{ 0, 1 \}^{|D_{\mathbf{M}}|} }{ x_{I}(\mathbf{c}) = \overline{x}_{I} \quad (\forall I) }.
\end{align}
Note that we do not need a one-hot encoding constraint such as Eq.~\eqref{eq:set-one-hot} in the binary system.

\subsection{\label{sec:formulation-feasible}Derivative structures}
Two distinct labelings may represent equivalent structures owing to the symmetry of the supercell.
For example, labelings $(0, 0, 1, 0, 1, 1)$ and $(0, 1, 1, 0, 0, 1)$ give equivalent structures in the example of Fig.~\ref{fig:terminology}.
Thus, we need to select a representative among the symmetrically equivalent labelings, which we call a nonequivalent labeling.
We refer to the corresponding substitutional structures as derivative structures.
We define the nonequivalent labeling as the minimum labeling in the lexicographic order among the equivalent labelings \cite{Hart2008,Mustapha2013}:
for example, labeling $(0, 0, 1, 0, 1, 1)$ is smaller than $(0, 1, 1, 0, 0, 1)$ in the lexicographic order because the first elements of both labelings are the same and the second element of $(0, 0, 1, 0, 1, 1)$ is smaller than that of $(0, 1, 1, 0, 0, 1)$.
We denote the set of lexicographically minimum labelings as $\tilde{\mathcal{C}}_{\mathrm{sym}}$.

We refer to a nonequivalent labeling $\tilde{\mathbf{c}}$ satisfying $x_{I, p}(\tilde{\mathbf{c}}) = \overline{x}_{I, p}$ for all atom types $p$ and sites $I$ as a feasible labeling.
The set of feasible labelings, $\tilde{\mathcal{C}}_{\mathrm{feasible}}$, is the intersection among the set of all possible labelings, the set of labelings with concentration $\overline{x}_{I, p}$, and the set of nonequivalent labelings,
\begin{align}
    \label{eq:set_derivative_structure}
    \tilde{\mathcal{C}}_{\mathrm{feasible}}
        = \tilde{\mathcal{C}}_{\mathrm{all}} \cap \tilde{\mathcal{C}}_{\mathrm{conc}} \cap \tilde{\mathcal{C}}_{\mathrm{sym}}.
\end{align}

In the binary system $(k=2)$, we refer to the set of nonequivalent reduced labelings as $\mathcal{C}_{\mathrm{sym}}$, which is also the set of lexicographically minimum reduced labelings.
A set of feasible reduced labelings is written as the intersection
\begin{align}
    \label{eq:set_derivative_structure_binary}
    \mathcal{C}_{\mathrm{feasible}}
        = \mathcal{C}_{\mathrm{conc}} \cap \mathcal{C}_{\mathrm{sym}}.
\end{align}

\subsection{\label{sec:formulation-corr}Correlation function and SQS}

SQSs \cite{SQS1,SQS2} are designed to exhibit the correlation functions of given clusters that are the closest to those of the perfectly random structure.
For the perfectly random structure, the correlation function can be determined by its concentration.
We denote the correlation function between atom types $p$ and $q$ for the perfectly random structure as
\begin{align}
    \overline{\Pi}_{ \alpha }^{pq} =
    \begin{cases}
        \overline{x}_{I, p} \overline{x}_{J, p} & (p = q) \\
        \overline{x}_{I, p} \overline{x}_{J, q} + \overline{x}_{I, q} \overline{x}_{J, p} & (p \neq q) \\
    \end{cases}
    \label{eq:correlation_function_random}
    ,
\end{align}
where the two point coordinates of $\alpha$ are equivalent to sites $I$ and $J$ in the primitive cell, respectively.

We give specific constraints for extracting SQSs from feasible labelings $\tilde{\mathcal{C}}_{\mathrm{feasible}}$.
An SQS is represented by a feasible labeling whose correlation functions are closest to the perfectly random structure.  %
We hierarchically search for feasible labelings whose correlation functions are the same as the perfectly random structure from the smallest pair cluster to a larger one because this hierarchical searching policy is compatible with a procedure for finding SQSs with the ZDD.
We define the set of labelings satisfying this constraint for pair cluster $\alpha$ as
\begin{align}
    \label{eq:set_corr}
    \tilde{\mathcal{C}}_{\alpha}
    = \bigcap_{p = 1}^{k} \bigcap_{q = 1}^{k} \tilde{\mathcal{C}}_{\alpha}^{pq},
\end{align}
where $\tilde{\mathcal{C}}_{\alpha}^{pq}$ is a set of labelings satisfying $\Pi_{ \alpha }^{p q} (\tilde{\mathbf{c}}) = \overline{\Pi}_{ \alpha }^{p q}$,
\begin{align}
    \tilde{\mathcal{C}}_{\alpha}^{pq}
    =
    \set{ \tilde{\mathbf{c}} \in \{ 0, 1 \}^{k |D_{\mathbf{M}}|} }{ \Pi_{ \alpha }^{pq} (\tilde{\mathbf{c}}) = \overline{\Pi}_{ \alpha }^{pq} }.
\end{align}

Finally, finding SQSs is equivalent to constructing the following set:
\begin{align}
    \label{eq:set_sqs}
    \tilde{\mathcal{C}}_{\mathrm{SQS}}^{N_{c}}
        = \tilde{\mathcal{C}}_{\mathrm{feasible}} \cap \left( \bigcap_{n=1}^{N_{c}} \tilde{\mathcal{C}}_{\alpha_{n}} \right),
\end{align}
where $\alpha_{n}$ indicates the $n$th NN pair cluster.
Each labeling in $\tilde{\mathcal{C}}_{\mathrm{SQS}}^{N_{c}}$ has the same correlation functions as the perfectly random structure up to the $N_{c}$th NN pair cluster.
In practice, the maximum number of distinct pair clusters, $N_{c}$, is adaptively incremented until immediately before $\tilde{\mathcal{C}}_{\mathrm{SQS}}^{N_{c}}$ becomes empty.

In the binary system, we write the correlation function for the perfectly random structure as $\overline{\Pi}_{\alpha}$.
We denote the set of feasible reduced labelings satisfying $\Pi_{ \alpha } (\mathbf{c}) = \overline{\Pi}_{ \alpha }$ as
\begin{align}
    \label{eq:set_corr_binary}
    \mathcal{C}_{\alpha}
        = \set{ \mathbf{c} \in \{ 0, 1 \}^{|D_{\mathbf{M}}|} }{ \Pi_{ \alpha } (\mathbf{c}) = \overline{\Pi}_{ \alpha } }.
\end{align}
The binary SQSs are equivalent to the intersection of Eqs.~\eqref{eq:set_derivative_structure_binary} and \eqref{eq:set_corr_binary},
\begin{align}
    \label{eq:set_sqs_binary}
    \mathcal{C}_{\mathrm{SQS}}^{N_{c}}
        = \mathcal{C}_{\mathrm{feasible}} \cap \left( \bigcap_{n=1}^{N_{c}} \mathcal{C}_{\alpha_{n}} \right).
\end{align}
For example, there are $|\mathcal{C}_{\mathrm{conc}}| = 20$ labelings with equiatomic concentration $AB$ for binary substitutional structures represented with the supercell shown in Fig.~\ref{fig:terminology}~(b).
When we consider equiatomic SQSs up to the first NN pair cluster $\alpha_{1}$, we search for reduced nonequivalent labelings $\mathbf{c}$ satisfying $\Pi_{\alpha_{1}}(\mathbf{c}) = \overline{\Pi}_{ \alpha_{1} } = 1/4$ out of $\mathcal{C}_{\mathrm{conc}}$.
There are $|\mathcal{C}_{\mathrm{feasible}}| = 3$ feasible reduced labelings, two of which satisfy $\Pi_{ \alpha_{1} } (\mathbf{c}) = \overline{\Pi}_{ \alpha_{1} }$; such labelings are expressed as
\begin{align*}
    \mathcal{C}_{\mathrm{SQS}}^{N_{c}=1}
        &= \mathcal{C}_{\mathrm{feasible}} \cap \mathcal{C}_{ \alpha_{1} } \\
        &= \left\{ (0, 0, 1, 0, 1, 1), (0, 0, 0, 1, 1, 1) \right\}.
\end{align*}

\section{\label{sec:algorithm}ZDD}
The number of feasible labelings increases exponentially with the index of supercell.
To deal with the exponential increase, we introduce an efficient data structure, the ZDD, in Sec.~\ref{sec:algorithm-zdd}.
We represent labelings with a fixed concentration using a ZDD in Sec.~\ref{sec:algorithm-zdd-concentration}.
We apply the procedure to represent nonequivalent labelings using the ZDD, which the authors proposed in Ref.~\onlinecite{doi:10.1063/5.0021663}.
The procedure is summarized in Sec.~\ref{sec:algorithm-zdd-nonequivalent}.
Finally, we introduce a procedure to construct a ZDD for representing labelings whose correlation functions are the same as those of the perfectly random structure in Sec.~\ref{sec:algorithm-zdd-sro}.

\subsection{\label{sec:algorithm-zdd}Relationship between binary decision tree and ZDD}

\begin{figure}[tb!]
    \centering
    \includegraphics[width=\linewidth]{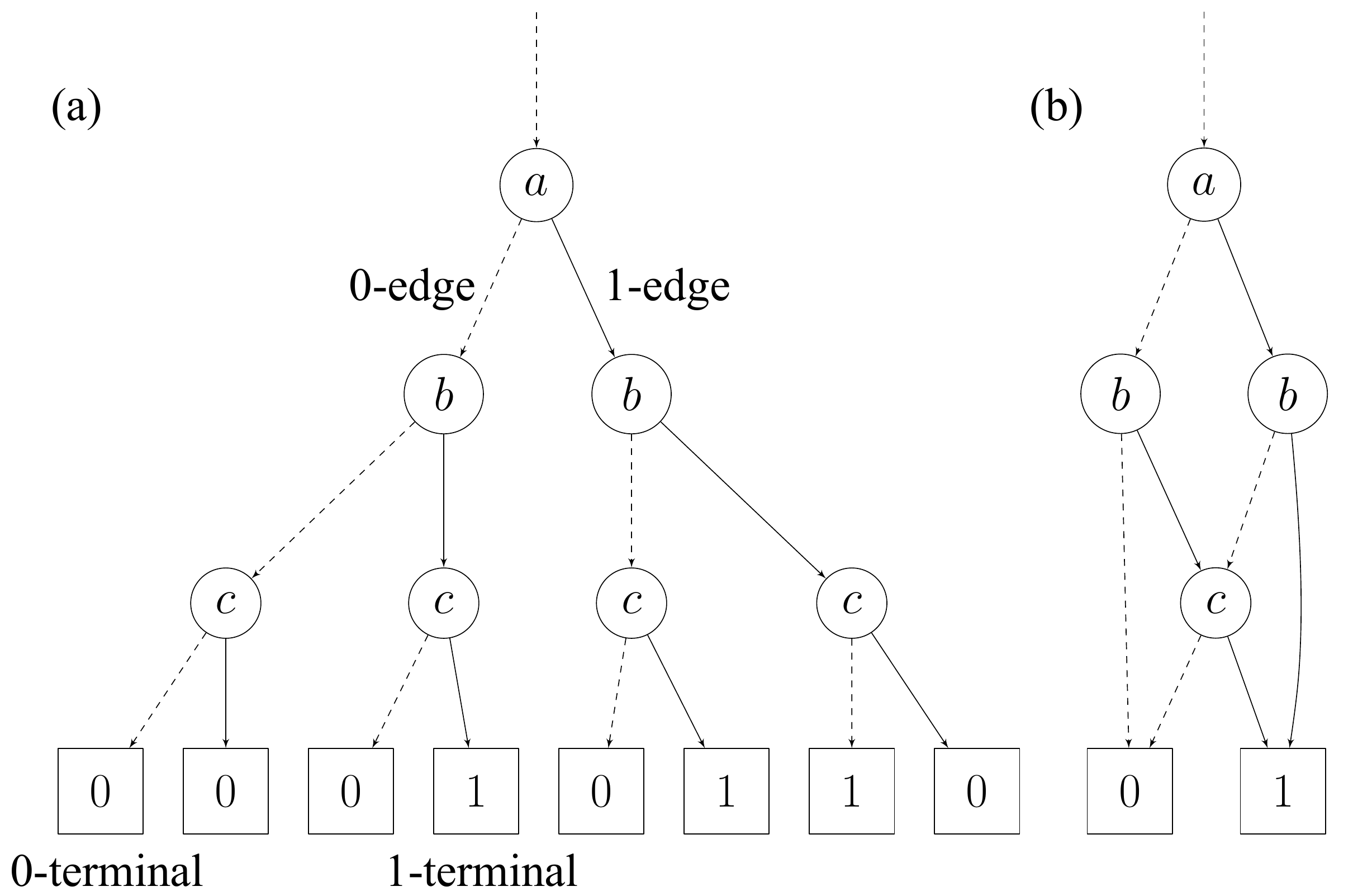}
    \caption{
        Binary decision tree and its ZDD.
        The solid and broken arrows indicate 1-edge and 0-edge, respectively.
        The square boxes with 1 and 0 indicate 1-terminal and 0-terminal nodes, respectively.
        (a) Binary decision tree representing the family of subsets $\{ \{a, b\}, \{a, c\}, \{b, c\} \}$.
        (b) ZDD of the binary decision tree derived by the reduction rules.
    }
    \label{fig:binary_tree_zdd}
\end{figure}

A binary decision tree \cite{knuth1978art} represents a family of subsets composed of $n$ elements satisfying given conditions.
By fixing the order of choosing each element, we can express the family of subsets as a tree structure.
The binary decision tree consists of terminal nodes, non-terminal nodes, and directed edges.
The terminal nodes are leaves of the binary decision tree.
The two kinds of terminal nodes, called the 1-terminal node and 0-terminal node, indicate whether or not the subset satisfies the given conditions, respectively.
Each non-terminal node, which corresponds to one of the $n$ elements, has two outgoing edges, the 1-edge and 0-edge.
These edges respectively indicate whether or not the corresponding element belongs to a subset.
Thus, a path from the root node to the 1-terminal node, called the 1-path, corresponds to a subset satisfying the given conditions.

A ZDD is one of the canonical and compact representations for a Boolean function \cite{Minato1993,sasao2014applications}.
A ZDD is derived by reducing a redundant part of a binary decision tree with the following two reduction rules:
\begin{itemize}
    \item (Node elimination) All non-terminal nodes whose 1-edges directly point to the 0-terminal nodes are deleted.
    \item (Node sharing) All equivalent nodes with the same child nodes and the same variable are shared.
\end{itemize}
The obtained irreducible ZDD is guaranteed to be canonical and independent of the order of applying the reduction rules.

Figure~\ref{fig:binary_tree_zdd}~(a) shows an example of a binary decision tree representing a family of subsets, $S = \{ \{a, b\}, \{a, c\}, \{b, c\} \}$, from three variables $a$, $b$, and $c$.
The three 1-paths in the binary decision tree correspond to the subsets in $S$, respectively.
Figure~\ref{fig:binary_tree_zdd}~(b) shows a ZDD derived from the binary decision tree shown in Fig.~\ref{fig:binary_tree_zdd}~(a).
The redundant non-terminal nodes are eliminated and merged by following the node elimination and node sharing rules.

Note that listing all the subsets is computationally heavier than constructing a ZDD.
Thus, if we construct a ZDD after listing all subsets, we lose efficiency.
Therefore, primitive set operations (e.g., intersection of sets) between two ZDDs \cite{bryant1986graph} and a frontier-based method \cite{sasao2014applications}, which directly derives a ZDD satisfying given constraints, have been used to improve the computational efficiency of deriving ZDDs in general.

\subsection{\label{sec:algorithm-zdd-concentration}Concentration}

\begin{figure*}[tb!]
    \centering
    \includegraphics[width=0.95\linewidth]{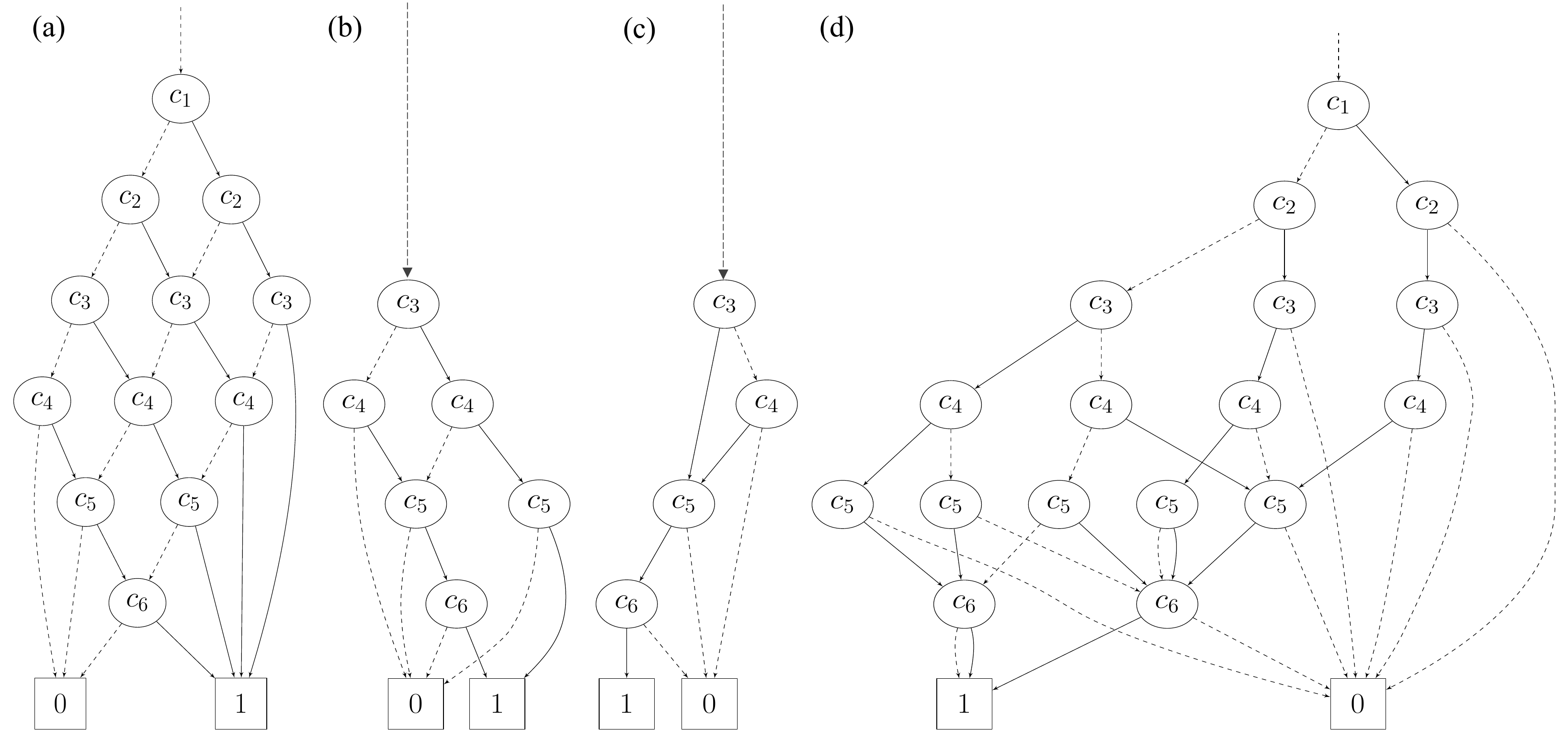}
    \caption{
        ZDDs of the two-dimensional example shown in Fig.~\ref{fig:terminology}.
        The non-terminal node $c_{i}$ corresponds to the occupation number of atom type 2 at site $i$ in the supercell.
        (a) ZDD representing reduced labelings of the two-dimensional example with equiatomic composition $\mathcal{C}_{\mathrm{conc}}$, which chooses three 1-edges from $c_{1}, \dots, c_{6}$.
        (b) ZDD representing feasible reduced labelings with equiatomic composition $\mathcal{C}_{\mathrm{feasible}}$.
        (c) ZDD representing SQSs up to the first NN pair cluster, $\mathcal{C}_{\mathrm{SQS}}^{N_{c}=1}$.
        (d) ZDD representing nonequivalent reduced labelings $\mathcal{C}_{\mathrm{sym}}$.
    }
    \label{fig:dd_example}
\end{figure*}

The ZDDs for $\tilde{\mathcal{C}}_{\mathrm{conc}}$ and $\mathcal{C}_{\mathrm{conc}}$ can be constructed with a procedure similar to that for constructing a ZDD representing labelings with a fixed concentration, as introduced in Ref.~\onlinecite{doi:10.1063/5.0021663}.
Figure~\ref{fig:dd_example}~(a) shows the ZDD corresponding to the binary equiatomic substitutional structures of Fig.~\ref{fig:terminology}~(b).
The constraint for the equiatomic composition is equivalent to choosing three 1-edges.
There are $\left( \begin{smallmatrix} 6 \\ 3 \end{smallmatrix} \right) = 20$ 1-paths and corresponding equiatomic substitutional structures.
In particular, the following 1-path in the ZDD corresponds to the derivative structure shown in Fig.~\ref{fig:terminology}~(c),
\begin{align*}
    c_{1}  \xrightarrow{ \mbox{\small{0-edge}} }
    &c_{2} \xrightarrow{ \mbox{\small{0-edge}} }
    c_{3}  \xrightarrow{ \mbox{\small{1-edge}} } \\
    &c_{4} \xrightarrow{ \mbox{\small{0-edge}} }
    c_{5}  \xrightarrow{ \mbox{\small{1-edge}} }
    c_{6}  \xrightarrow{ \mbox{\small{1-edge}} }
    \fbox{1}.
\end{align*}

\subsection{\label{sec:algorithm-zdd-nonequivalent}Nonequivalent labelings}

Recently, the authors have proposed the frontier-based ZDD method for efficiently enumerating derivative structures \cite{doi:10.1063/5.0021663}.
The algorithm for eliminating symmetrically equivalent labelings is based on Ref.~\onlinecite{Horiyama2018}, in which the enumeration of all non-isomorphic subgraphs of a given graph up to the automorphism was proposed.

Figure~\ref{fig:dd_example}~(b) shows the ZDD corresponding to the binary equiatomic derivative structures for the supercell shown in Fig.~\ref{fig:terminology}~(b).
The derivative structure in Fig.~\ref{fig:terminology}~(c) corresponds to the following 1-path in the ZDD:
\begin{align*}
    c_{3} \xrightarrow{ \mbox{\small{1-edge}} }
    c_{4} \xrightarrow{ \mbox{\small{0-edge}} }
    c_{5} \xrightarrow{ \mbox{\small{1-edge}} }
    c_{6} \xrightarrow{ \mbox{\small{1-edge}} }
    \fbox{1}.
\end{align*}
In this 1-path, two non-terminal nodes, $c_{1}$ and $c_{2}$, are deleted by the node elimination rule, which indicates $c_{1} = c_{2} = 0$.
If $c_{1}$ or $c_{2}$ is assigned to one, the constraint for $\mathcal{C}_{\mathrm{sym}}$ is never satisfied with the equiatomic composition.

Note that the order of intersections in constructing a ZDD strongly affects the total computational time and required memory.
In the two-dimensional example, we can construct $\mathcal{C}_{\mathrm{feasible}}$ shown in Fig.~\ref{fig:dd_example}~(b) by extracting nonequivalent labelings from $\mathcal{C}_{\mathrm{conc}}$ shown in Fig.~\ref{fig:dd_example}~(a) or by extracting labelings with a fixed concentration from $\mathcal{C}_{\mathrm{sym}}$ shown in Fig.~\ref{fig:dd_example}~(d).
We empirically observe that the former procedure is more efficient than the latter, which corresponds to the fact that the ZDD of $\mathcal{C}_{\mathrm{conc}}$ has fewer non-terminal nodes than that of $\mathcal{C}_{\mathrm{sym}}$.

\subsection{\label{sec:algorithm-zdd-sro}Correlation functions}

\begin{figure}[tb!]
    \centering
    \includegraphics[width=\linewidth]{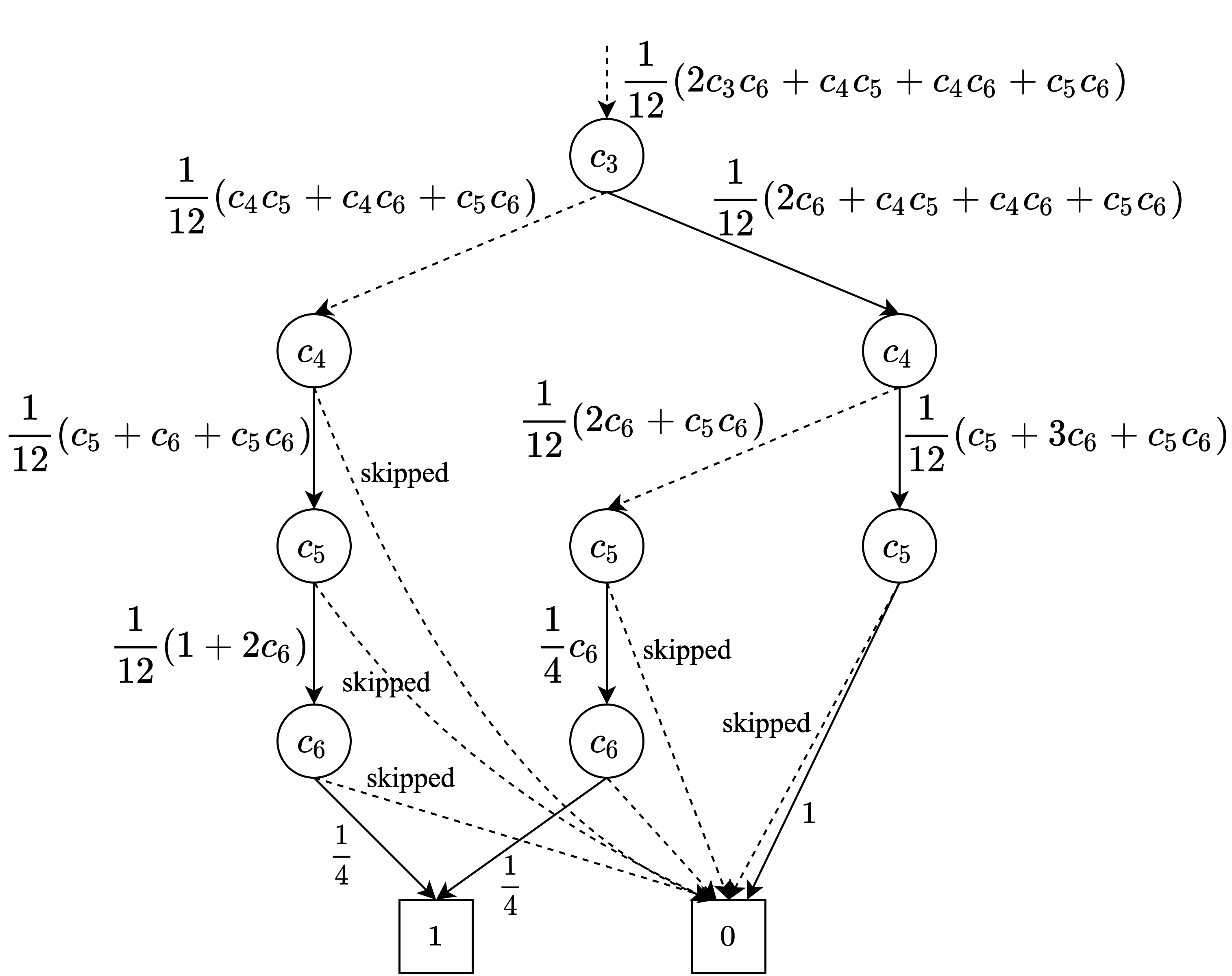}
    \caption{
        Development of a ZDD for $\Pi_{ \alpha_{1} }(\mathbf{c}) = \overline{\Pi}_{ \alpha_{1} } = 1/4$ of the two-dimensional example.
        A partially determined value of the correlation function $\Pi_{ \alpha_{1} }(\mathbf{c})$ is attached to each edge from non-terminal nodes.
        An edge with ``skipped" goes to the 0-terminal node because the corresponding labeling is not a feasible labeling.
    }
    \label{fig:dd_sqs_frontiermethod}
\end{figure}

After the ZDD for $\tilde{\mathcal{C}}_{\mathrm{feasible}}$ is constructed, we take the intersection between it and a ZDD to represent labelings satisfying the constraint $\Pi_{\alpha}^{pq}(\tilde{\mathbf{c}}) = \overline{\Pi}_{\alpha}^{pq}$, $\tilde{\mathcal{C}_{\alpha}^{pq}}$.
This constraint is regarded as a quadratic equation of the labeling $\tilde{\mathbf{c}}$.
Because the left-hand side of the equation monotonically increases with each label $\tilde{c}_{i}$, we can eliminate a partially determined labeling that cannot satisfy the constraint.
Moreover, two nodes in a ZDD are merged when partially determined terms of $\Pi_{ \alpha }^{pq}(\tilde{\mathbf{c}})$ are equal.
This ZDD construction for SQSs avoids explicitly listing all labelings and checking their correlation functions one by one because the correlation functions are efficiently examined on $\tilde{\mathcal{C}}_{\mathrm{feasible}}$.
Although the ZDDs for correlation functions $\tilde{\mathcal{C}}_{\alpha}$ and $\mathcal{C}_{\alpha}$ are represented by a compressed form of a corresponding binary decision tree, their construction is computationally more expensive than the other ZDDs.
Thus, we impose the constraints in terms of correlation functions after constructing the set of feasible labelings $\tilde{\mathcal{C}}_{\mathrm{feasible}}$ and $\mathcal{C}_{\mathrm{feasible}}$.

In a binary system, the ZDD for the constraint $\Pi_{ \alpha }(\mathbf{c}) = \overline{\Pi}_{ \alpha }$ can be constructed as well as the multicomponent systems.
Figure~\ref{fig:dd_sqs_frontiermethod} shows the development of the ZDD $\mathcal{C}_{\mathrm{feasible}} \cap \mathcal{C}_{ \alpha_{1} }$ for the two-dimensional example with the first NN pair cluster $\alpha_{1}$.
The equation attached to each edge in $\mathcal{C}_{\mathrm{feasible}} \cap \mathcal{C}_{ \alpha_{1} }$ corresponds to a partially determined value of the correlation function for $\alpha_{1}$.
Figure~\ref{fig:dd_example}~(c) shows the irreducible ZDD for the two-dimensional example further compressed by the two reduction rules.

\section{\label{sec:results}Results and Discussion}

\subsection{\label{sec:results-pareto-optimal-sqs}Pareto-optimal SQS}

\begin{table}[tb]
    \centering
    \caption{
        Numbers of labelings in search for binary fcc-based SQSs of $AB$ with 40 sites with transformation matrix $\mathbf{M} = [[1, 0, 0], [1, 2, 0], [5, 0, 20]]$.
    }
    \label{tab:number_into_fcc_binary_sqs}
    \setlength\tabcolsep{6pt}  %
    \begin{ruledtabular}
    \begin{tabular}{lr}
        Constraints & Number of labelings \\
        \hline
        No constraint & 1,099,511,627,776 \\
        + Concentration $AB$ & 137,846,528,820 \\
        + Nonequivalent & 863,005,322 \\
        + Up to the first NN & 77,521,770 \\
        + Up to the second NN & 8,564,691 \\
        + Up to the third NN & 564,221 \\
        + Up to the fourth NN & 27,192 \\
        + Up to the fifth NN & 2,773 \\
        + Up to the sixth NN & 770 \\
        + Up to the seventh NN & 68 \\
        + Up to the eighth NN & 68 \\
        + Up to the ninth NN & 12 \\
        + Up to the tenth NN & 12 \\
    \end{tabular}
    \end{ruledtabular}
\end{table}

\begin{table*}[tb]
    \centering
    \caption{
        Pareto-optimal fcc-based SQSs for $AB$, $A_{2}B$, $A_{3}B$, $A_{3}B_{2}$, $A_{4}B$, $A_{5}B$, $ABC$, $A_{2}BC$, and $ABCD$.
        The second column shows the number of sites in SQSs.
        The norms of the differences in correlation functions $||\Pi_{\alpha}(\tilde{\mathbf{c}}) - \overline{\Pi}_{\alpha}^{pq}||$ for pair clusters up to the tenth NN are also shown.
        For ternary and quaternary systems, the averages of the norms over atom types, $p$ and $q$, are shown.
        The last column shows the differences in correlation functions of the smallest triplet.
        The information of the pair clusters is shown in Appendix~\ref{sec:app-table}.
    }
    \label{tab:result_1_sqs_fcc}
    \begin{ruledtabular}
    \begin{tabular}{ll|cccccccccc|c}
        &  & \multicolumn{10}{c|}{Pair clusters} & Triplet cluster \\
        & SQS-$N$ & First & Second & Third & Fourth & Fifth & Sixth & Seventh & Eighth & Ninth & Tenth & Nearest \\
        \hline
        $AB$ & SQS-2 & \textbf{0} & 0.25 & 0 & 0.25 & 0 & 0.25 & 0 & 0.25 & 0 & 0 & 0 \\
        & SQS-8 & \textbf{0} & \textbf{0} & 0.042 & 0.083 & 0 & 0 & 0.042 & 0.167 & 0 & 0 & 0 \\
        & SQS-16 & \textbf{0} & \textbf{0} & \textbf{0} & \textbf{0} & \textbf{0} & \textbf{0} & \textbf{0} & 0.25 & 0 & 0 & 0 \\
        & SQS-32 & \textbf{0} & \textbf{0} & \textbf{0} & \textbf{0} & \textbf{0} & \textbf{0} & \textbf{0} & \textbf{0} & 0.021 & 0.042 & 0 \\
        & SQS-40 & \textbf{0} & \textbf{0} & \textbf{0} & \textbf{0} & \textbf{0} & \textbf{0} & \textbf{0} & \textbf{0} & \textbf{0} & \textbf{0} & 0 \\
        \hline
        $A_{2}B$ & SQS-3 & \textbf{0} & 0.111 & 0.111 & 0 & 0.111 & 0.111 & 0 & 0.111 & 0.111 & 0.222 & 0.037 \\
        & SQS-9 & \textbf{0} & \textbf{0} & 0.028 & 0 & 0.037 & 0.028 & 0.019 & 0.037 & 0.037 & 0.074 & 0.009 \\
        & SQS-12 & \textbf{0} & \textbf{0} & \textbf{0} & 0.014 & 0.014 & 0.181 & 0.014 & 0.167 & 0 & 0 & 0.005 \\
        & SQS-18 & \textbf{0} & \textbf{0} & \textbf{0} & \textbf{0} & 0.023 & 0.028 & 0.002 & 0.093 & 0 & 0.046 & 0.009 \\
        & SQS-21 & \textbf{0} & \textbf{0} & \textbf{0} & \textbf{0} & \textbf{0} & 0.044 & 0.008 & 0.016 & 0 & 0 & 0.005 \\
        & SQS-27 & \textbf{0} & \textbf{0} & \textbf{0} & \textbf{0} & \textbf{0} & \textbf{0} & \textbf{0} & \textbf{0} & \textbf{0} & 0.056 & 0 \\
        \hline
        $A_{3}B$ & SQS-8 & \textbf{0} & 0.104 & 0 & 0.062 & 0.042 & 0.062 & 0.042 & 0.021 & 0 & 0 & 0.016 \\
        & SQS-16 & \textbf{0} & \textbf{0} & \textbf{0} & 0.021 & 0.01 & 0.016 & 0.021 & 0.062 & 0.026 & 0.01 & 0 \\
        & SQS-32 & \textbf{0} & \textbf{0} & \textbf{0} & \textbf{0} & \textbf{0} & \textbf{0} & \textbf{0} & 0.021 & 0.021 & 0.021 & 0 \\
        \hline
        $A_{3}B_{2}$ & SQS-25 & \textbf{0} & \textbf{0} & \textbf{0} & \textbf{0} & \textbf{0} & 0.02 & 0.015 & 0.04 & 0.013 & 0.027 & 0.001 \\
        \hline
        $A_{4}B$ & SQS-25 & \textbf{0} & \textbf{0} & \textbf{0} & \textbf{0} & \textbf{0} & \textbf{0} & 0.002 & 0.027 & 0 & 0.027 & 0.003 \\
        & SQS-50 & \textbf{0} & \textbf{0} & \textbf{0} & \textbf{0} & \textbf{0} & \textbf{0} & \textbf{0} & \textbf{0} & \textbf{0} & 0.02 & 0.001 \\
        \hline
        $A_{5}B$ & SQS-6 & \textbf{0} & 0.028 & 0 & 0 & 0.028 & 0.056 & 0.028 & 0.028 & 0 & 0.028 & 0.005 \\
        & SQS-12 & \textbf{0} & \textbf{0} & \textbf{0} & \textbf{0} & 0.014 & 0.014 & 0.007 & 0.028 & 0.014 & 0.097 & 0.005 \\
        & SQS-36 & \textbf{0} & \textbf{0} & \textbf{0} & \textbf{0} & \textbf{0} & \textbf{0} & 0.005 & 0.009 & 0.016 & 0.019 & 0.001 \\
        \hline
        $ABC$ & SQS-3 & \textbf{0} & 0.064 & 0.064 & 0 & 0.064 & 0.064 & 0 & 0.064 & 0.064 & 0.128 & 0.222 \\
        & SQS-9 & \textbf{0} & \textbf{0} & 0.016 & 0 & 0 & 0.016 & 0 & 0 & 0.016 & 0.128 & 0.056 \\
        & SQS-12 & \textbf{0} & \textbf{0} & \textbf{0} & 0.013 & 0.013 & 0.099 & 0.008 & 0.091 & 0.016 & 0 & 0.028 \\
        & SQS-24 & \textbf{0} & \textbf{0} & \textbf{0} & \textbf{0} & \textbf{0} & 0.016 & 0 & 0.064 & 0 & 0 & 0.003 \\
        & SQS-27 & \textbf{0} & \textbf{0} & \textbf{0} & \textbf{0} & \textbf{0} & \textbf{0} & \textbf{0} & \textbf{0} & \textbf{0} & 0.032 & 0 \\
        \hline
        $A_{2}BC$ & SQS-8 & \textbf{0} & 0.014 & 0.023 & 0.012 & 0.023 & 0.021 & 0.006 & 0.059 & 0.01 & 0.014 & 0.125 \\
        & SQS-16 & \textbf{0} & \textbf{0} & 0.007 & 0.042 & 0.007 & 0 & 0 & 0.057 & 0 & 0.014 & 0 \\
        \hline
        $ABCD$ & SQS-8 & \textbf{0} & 0.014 & 0.023 & 0.012 & 0.023 & 0.021 & 0.006 & 0.059 & 0.01 & 0.014 & 0.125 \\
        & SQS-16 & \textbf{0} & \textbf{0} & 0.007 & 0.042 & 0.007 & 0 & 0 & 0.057 & 0 & 0.014 & 0 \\
    \end{tabular}
    \end{ruledtabular}
\end{table*}

\begingroup
\squeezetable
\begin{table*}
    \centering
    \caption{
        Pareto-optimal hcp-based SQSs for compositions $AB$, $ABC$, and $ABCD$.
        The information of the pair clusters is shown in Appendix~\ref{sec:app-table}.
    }
    \label{tab:result_1_sqs_hcp}
    \begin{ruledtabular}
    \begin{tabular}{ll|cccccccccccccccccc|c}
        &  & \multicolumn{18}{c|}{Pair clusters} & Triplet cluster \\
        & SQS-$N$ & 1 & 2 & 3 & 4 & 5 & 6 & 7 & 8 & 9 & 10 & 11 & 12 & 13 & 14 & 15 & 16 & 17 & 18 & Nearest \\
        \hline
        $AB$ & SQS-4 & \textbf{0} & 0.083 & 0 & 0.25 & 0.083 & 0 & 0.083 & 0.25 & 0 & 0.083 & 0 & 0 & 0.25 & 0 & 0.083 & 0 & 0 & 0.083 & 0.061 \\
        & SQS-8 & \textbf{0} & \textbf{0} & \textbf{0} & \textbf{0} & 0.083 & 0 & 0.083 & 0.083 & 0 & 0 & 0 & 0 & 0.167 & 0 & 0.083 & 0 & 0 & 0 & 0.014 \\
        & SQS-16 & \textbf{0} & \textbf{0} & \textbf{0} & \textbf{0} & \textbf{0} & \textbf{0} & \textbf{0} & 0.25 & 0 & 0 & 0 & 0 & 0 & 0 & 0 & 0 & 0 & 0 & 0 \\
        & SQS-24 & \textbf{0} & \textbf{0} & \textbf{0} & \textbf{0} & \textbf{0} & \textbf{0} & \textbf{0} & \textbf{0} & \textbf{0} & 0.014 & 0 & 0 & 0.097 & 0 & 0.014 & 0 & 0 & 0.056 & 0 \\
        & SQS-32 & \textbf{0} & \textbf{0} & \textbf{0} & \textbf{0} & \textbf{0} & \textbf{0} & \textbf{0} & \textbf{0} & \textbf{0} & \textbf{0} & \textbf{0} & \textbf{0} & \textbf{0} & \textbf{0} & \textbf{0} & \textbf{0} & \textbf{0} & \textbf{0} & 0 \\
        \hline
        $ABC$ & SQS-6 & \textbf{0} & \textbf{0} & 0.032 & 0.064 & 0.064 & 0.032 & 0 & 0 & 0.032 & 0.032 & 0.032 & 0 & 0 & 0 & 0 & 0.064 & 0.064 & 0.128 & 0.111 \\
        & SQS-12 & \textbf{0} & \textbf{0} & \textbf{0} & 0.064 & 0 & 0 & 0.032 & 0.064 & 0 & 0.064 & 0 & 0 & 0.032 & 0 & 0 & 0 & 0 & 0 & 0.105 \\
        & SQS-18 & \textbf{0} & \textbf{0} & \textbf{0} & \textbf{0} & 0.043 & 0 & 0.011 & 0.043 & 0 & 0.005 & 0 & 0 & 0.032 & 0 & 0.011 & 0 & 0 & 0.064 & 0.105 \\
        \hline
        $ABCD$ & SQS-16 & \textbf{0} & \textbf{0} & \textbf{0} & 0.029 & 0.014 & 0 & 0.014 & 0.017 & 0 & 0.01 & 0 & 0 & 0.029 & 0 & 0.014 & 0 & 0 & 0 & 0.036 \\
    \end{tabular}
    \end{ruledtabular}
\end{table*}
\endgroup

\begin{figure}
    \centering
    \includegraphics[width=0.8\linewidth]{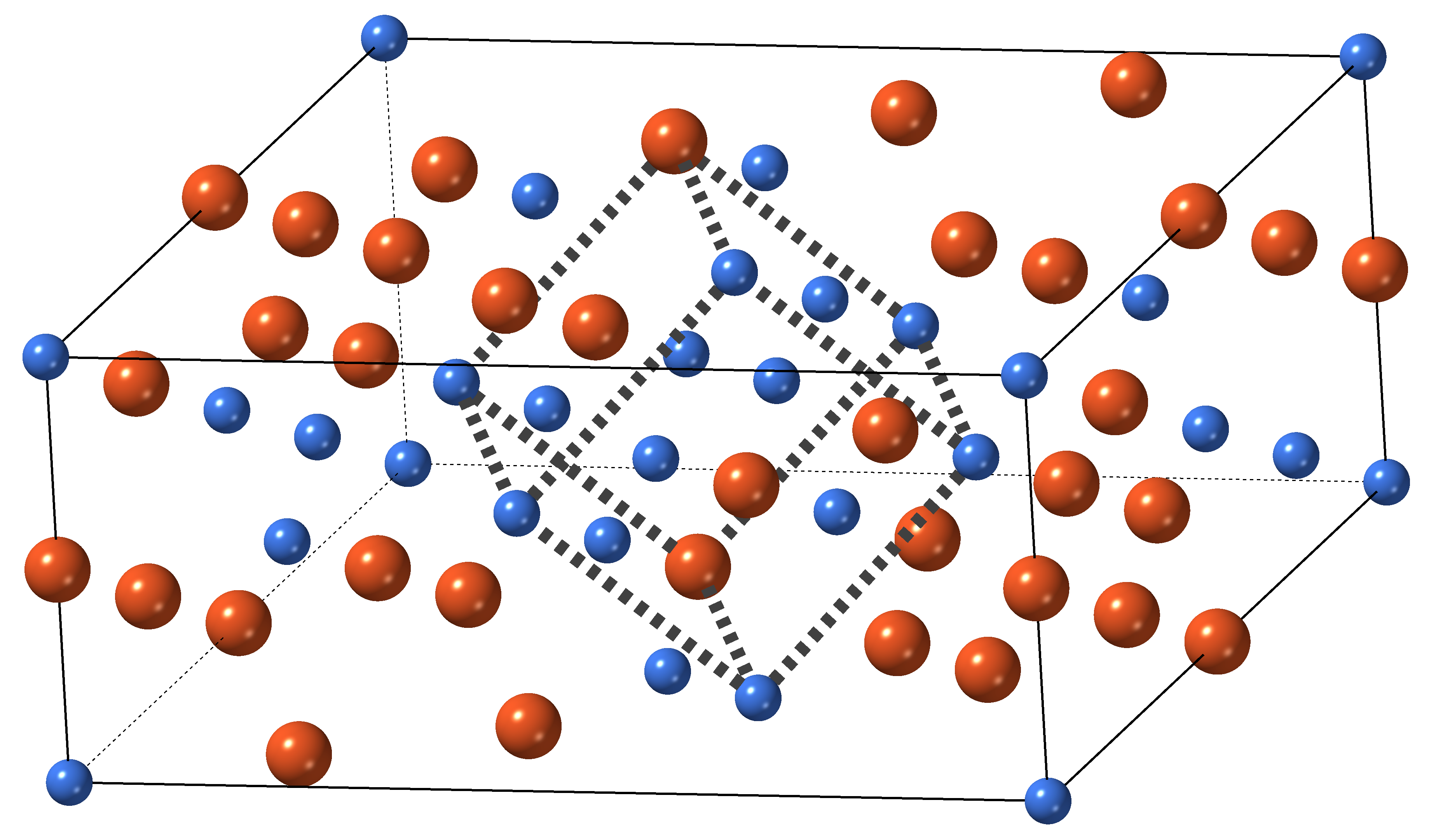}
    \caption{
        Crystal structure of a binary fcc-based SQS-40 of $AB$.
        The orange and blue balls represent atom types $A$ and $B$, respectively.
        The tilted cube with the dotted lines indicates a conventional fcc unit cell.
    }
    \label{fig:sqs_fcc_binary}
\end{figure}

We demonstrate applications of the present ZDD-based method for fcc-based and hcp-based SQSs.
We show the computational details of the present ZDD-based method.
We use \textsc{spglib}~\cite{spglib} and \textsc{pymatgen}~\cite{pymatgen} to obtain symmetrically nonequivalent pair clusters.
We choose the $c/a$ ratio to be ideal for the hcp primitive cell.
We use \textsc{tdzdd}~\cite{Iwashita13,TdZdd} to implement the frontier-based algorithm for constructing ZDDs.

We present an example of finding binary equiatomic fcc-based SQSs with 40 sites (SQS-40).
If we do not use a ZDD and explicitly tabulate labelings when imposing constraints, we have to store $2^{40} (\sim 10^{12})$ labelings.
On the other hand, the ZDD can store all the labelings in a compressed manner.
First, we enumerate all possible supercells of the fcc structure with 40 sites by enumerating all nonequivalent Hermite normal forms whose index is 40 \cite{Hart2008}.
We obtain 286 nonequivalent transformation matrices and construct $\mathcal{C}_{\mathrm{SQS}}^{N_{c}}$ for every transformation matrix.
The process of developing $\mathcal{C}_{\mathrm{SQS}}^{N_{c}}$ for transformation matrix
\begin{align}
    \label{eq:transformation-matrix-40}
    \mathbf{M} =
        \begin{pmatrix}
            1 & 0 & 0 \\
            1 & 2 & 0 \\
            5 & 0 & 20 \\
        \end{pmatrix}
\end{align}
is shown in Table~\ref{tab:number_into_fcc_binary_sqs}.
We construct the ZDD for representing 137,846,528,820 equiatomic labelings, $\mathcal{C}_{\mathrm{conc}}$, and extract 863,005,322 feasible labelings $\mathcal{C}_{\mathrm{feasible}}$.
Since fcc is a highly symmetric structure, the constraint for eliminating equivalent labelings significantly reduces the number of labelings by a factor of about 100.
We then extract feasible labelings that are the same as the perfectly random structure in terms of the first NN pair cluster by taking the intersection between $\mathcal{C}_{\mathrm{feasible}}$ and $\mathcal{C}_{\alpha_{1}}$.
Similarly, we extract feasible labelings that are the same as the perfectly random structure in terms of up to the tenth NN pairs by successively taking the intersection between $\mathcal{C}_{\mathrm{SQS}}^{N_{c}-1}$ and $\mathcal{C}_{\alpha_{N_{c}}}$.
The number of labelings gradually decreases from $|\mathcal{C}_{\mathrm{SQS}}^{N_{c}=1}| = 77,521,770$ to $|\mathcal{C}_{\mathrm{SQS}}^{N_{c}=10}| = 12$.

By developing $\mathcal{C}_{\mathrm{SQS}}^{N_{c}}$ for other transformation matrices with the index of 40, it is revealed that we can find labelings that are the same as the perfectly random structure in terms of all pairs up to the tenth NN.
Thus, the final 12 structures with the transformation matrix of Eq.~\eqref{eq:transformation-matrix-40} are obtained as SQS-40 up to the tenth NN pair, and the upper bound of $N_{c}$ not giving the empty $\mathcal{C}_{\mathrm{SQS}}^{N_{c}}$, $N_{c}^{\max}$, is ten for SQS-40.
If multiple labelings are included in the final set of labelings, we pick up one of the labelings that is the closest to the perfectly random structure in terms of additional triplets or quadruples.
The correlation functions for the additional clusters can be calculated by a general procedure without using the ZDD.

We search for SQSs up to the limit of the index imposed by the computational resources
\footnote{
    We used a workstation powered by Intel\textregistered Xeon\textregistered Gold 6230 (2.1~GHz) with 3072~GB RAM.
}.
For fcc, the present ZDD-based method exhaustively searches for SQSs with up to 40, 45, 40, 35, 50, 60, 27, 24, and 20 sites for compositions $AB$, $A_{2}B$, $A_{3}B$, $A_{3}B_{2}$, $A_{4}B$, $A_{5}B$, $ABC$, $A_{2}BC$, and $ABCD$, respectively.
At the same time, the SQSs with the maximum index are not necessarily optimal because $N_{c}^{\max}$ does not monotonically increase with the index of a supercell.
For example, $N_{c}^{\max}$ is obtained to be three for SQS-45 in $A_{2}B$, which is smaller than that for SQS-27.
Thus, we select Pareto-optimal SQSs from the entire set of SQSs, which are determined from the trade-off relationship between the index and $N_{c}^{\max}$.
The obtained fcc-based Pareto-optimal SQSs are summarized in Table~\ref{tab:result_1_sqs_fcc}.
If multiple SQSs are found for an index, we tabulate an SQS whose correlation function of the smallest triplet cluster is closest to the perfectly random structure, although different choices of additional clusters are also possible.
These Pareto-optimal SQSs are presented in Supplemental Material \cite{SI}.

The fcc-based SQS-2 and SQS-8 derived by the ZDD-based method for composition $AB$ were reported in Refs.~\onlinecite{SQS1,SQS2}.
The ZDD-based method newly finds SQS-32 and SQS-40 with $N_{c}^{\max}=8 \;\mbox{and}\; 10$, respectively.
For illustration, one of the obtained SQS-40 for fcc $AB$ is shown in Fig.~\ref{fig:sqs_fcc_binary}.
Similarly, the ZDD-based method newly finds SQS-27 with $N_{c}^{\max}=9$ for composition $ABC$, whereas an fcc-based SQS with 18 sites and $N_{c}^{\max}=2$ was reported in Ref.~\onlinecite{PhysRevB.76.144204}.
Reference~\onlinecite{PhysRevB.76.144204} also showed stochastically searched $ABC$ structures with 24, 36, and 48 sites with $N_{c}^{\max}=3$.
Note that it is difficult to fairly compare the current SQSs with the previous ones because they are generated by using different procedures and definitions of the similarity to the perfectly random structure.
Nevertheless, the newly found SQS-27 has a larger $N_{c}^{\max}$ and is represented by a smaller number of sites than the previously reported stochastic ones.

The present ZDD-based method exhaustively searches for hcp-based SQSs with up to 42, 27, and 20 sites for $AB$, $ABC$, and $ABCD$, respectively.
The hcp-based Pareto-optimal SQSs are summarized in Table~\ref{tab:result_1_sqs_hcp}.
The hcp-based SQS-8 and SQS-16 for $AB$, which are reproduced by the ZDD-based method, were already reported in Ref.~\onlinecite{SQS_hcp}.
The ZDD-based method searches for SQS-24 and SQS-32 with $N_{c}^{\max} = 9 \;\mbox{and}\; 18$, respectively, which were not previously reported.
The ZDD-based method also finds SQS-18 with $N_{c}^{\max}=4$ for composition $ABC$, whereas \textsc{atat} provides an SQS-48 with $N_{c}^{\max}=2$ searched for by a stochastic method \cite{sqs2tdb}.

\subsection{\label{sec:results-computational-performance}Computational performance}

\begin{figure}[tb]
    \centering
    \includegraphics[width=\linewidth]{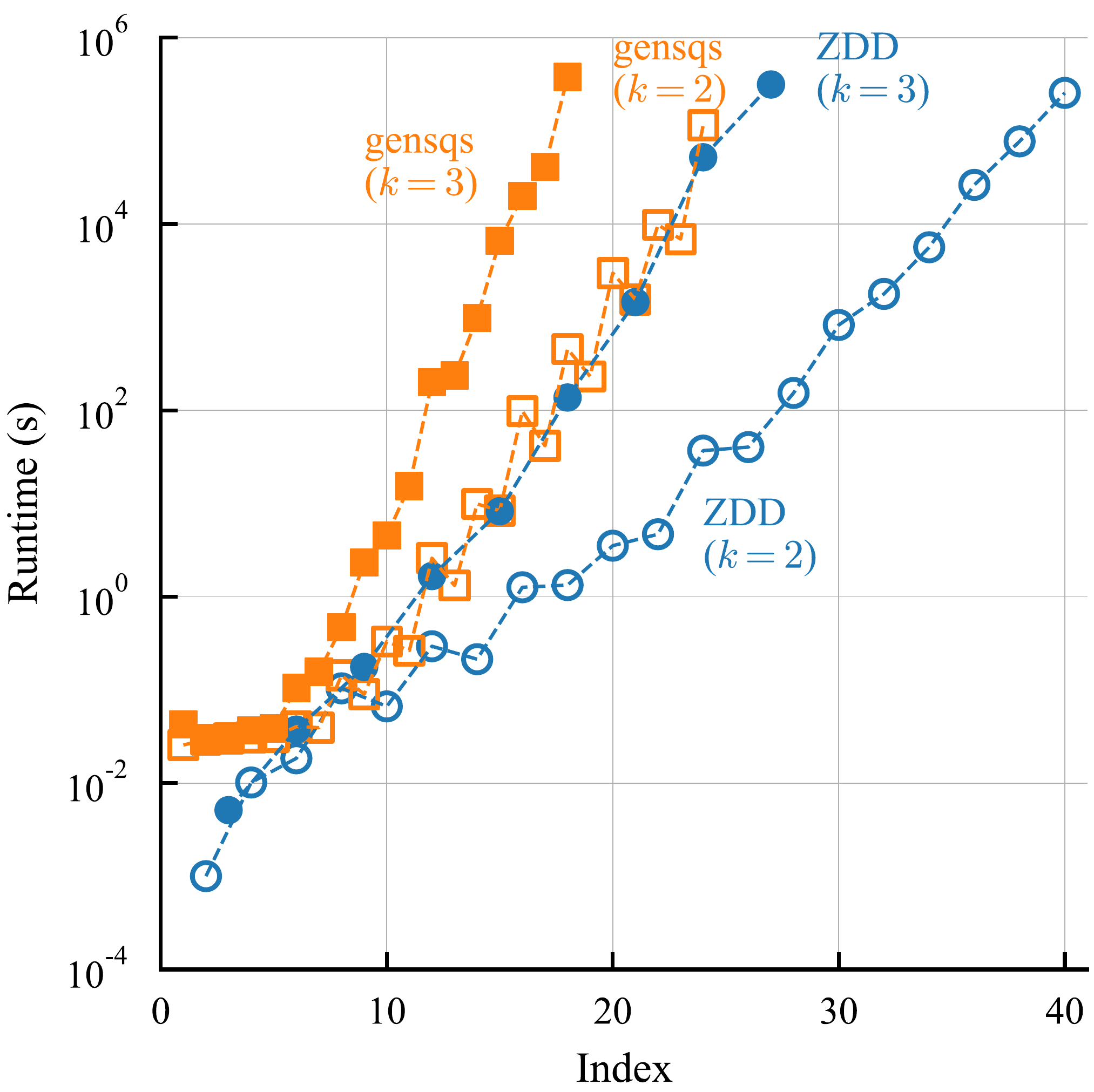}
    \caption{
        Computational times to search for equiatomic fcc-based SQSs with the present ZDD-based method and the previous enumeration method on a logarithmic scale.
        The blue open and closed circles stand for the runtimes for constructing ZDDs for binary and ternary systems, respectively.
        The orange open and closed squares stand for the runtimes for naively searching for SQSs with the previous method.
        The horizontal axis indicates the index of the supercell.
    }
    \label{fig:sqs_runtime}
\end{figure}

We compare the performance of the present ZDD-based method with that of the previous enumeration method implemented in \textsc{gensqs} in \textsc{atat} \cite{atat}.
The previous method searches for SQSs by explicitly listing all nonequivalent derivative structures with a given index and checking their correlation functions one by one.
We measure the runtime of both methods for searching for binary and ternary fcc-based SQSs of $AB$ and $ABC$.
Unlike the ZDD-based method, the previous method needs to fix the range of considered pair clusters beforehand to define the similarity to the perfectly random structure.
For a binary system, we limit the range of pair clusters up to the eighth one.
For a ternary system, we limit the range of pair clusters up to the fifth one.
Although the runtime slightly depends on the choice of the clusters, the following comparison is still useful.
The results of these calculations are shown in Fig.~\ref{fig:sqs_runtime}.
The ZDD-based method improves the base of the exponential runtime to around half that of the previous enumeration method: hence, the ZDD-based method can find SQSs much faster than the previous method.
For example, the runtime for searching for binary SQS-24 with the ZDD is about 3000 times shorter than the previous enumeration method, and the improvement in runtime further increases with a larger index.
The improvement in runtime and the decrease in the required amount of memory enable SQSs with larger indices to be found.
Thus, the ZDD-based method for searching for SQSs is much more efficient than the previous enumeration method.
The stochastic approach for finding SQS is also implemented in \textsc{atat}.
In Appendix~\ref{sec:app-mcsqs}, we compare the performance of the present ZDD-based method with that of the stochastic method in conjunction with describing their methodological differences.

Finally, we mention the limitations of the current ZDD-based method.
Firstly, as described above, the current ZDD-based method finds fcc-based Pareto-optimal SQSs with up to 16 sites for compositions $A_{2}BC$ and $ABCD$.
However, they reproduce the correlation functions of the perfectly random structure only for up to the second NN pairs.
Thus, the current ZDD-based method is insufficient for finding larger SQSs that reproduce the correlation functions of larger clusters.
In such a case, a stochastic method can be a practical solution.
Secondly, the current ZDD-based method is formulated using the correlation functions of pair clusters.
Although an extension of the formulation to triplet and quadruple clusters is possible in a straightforward manner, it causes a drastic increase in the ZDD size.
In practice, after the number of candidate structures is significantly reduced by the ZDD-based method for pair clusters, their correlation functions of triplet and quadruple clusters can be easily calculated using a general procedure without the ZDD.
Similarly, it is possible to replace the conditions that the correlation functions of the SQS are exactly the same as those of the perfectly random structure with relaxed conditions given by thresholds of the correlation functions.
However, the relaxation of the conditions should also increase the ZDD size.
Therefore, it is computationally challenging to apply such relaxation to a ZDD composed of many structures.

\subsection{\label{sec:results-realistic-systems}Application to realistic systems}

\begin{figure}[tb]
    \centering
    \includegraphics[width=\linewidth]{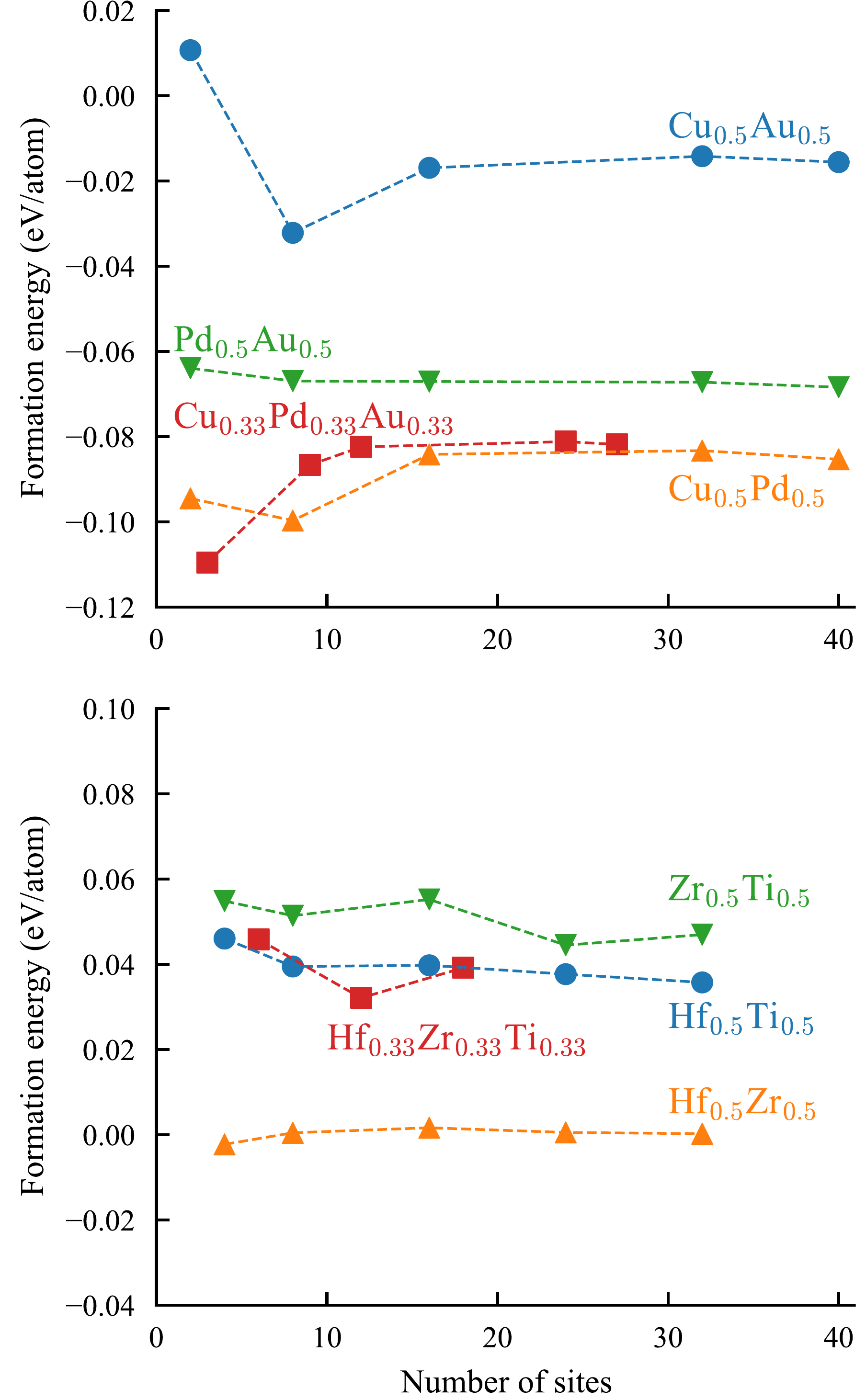}
    \caption{
        Formation energies of SQSs.
        The horizontal axis indicates the number of sites in SQSs.
        The upper figure shows the results for fcc random structures.
        The blue circles, orange up triangles, green down triangles, and red squares indicate Cu$_{0.5}$Au$_{0.5}$, Cu$_{0.5}$Pd$_{0.5}$, Pd$_{0.5}$Au$_{0.5}$, and Cu$_{0.33}$Pd$_{0.33}$Au$_{0.33}$, respectively.
        The lower figure shows the results for hcp random structures.
        The blue circles, orange up triangles, green down triangles, and red squares indicate Hf$_{0.5}$Ti$_{0.5}$, Hf$_{0.5}$Zr$_{0.5}$, Zr$_{0.5}$Ti$_{0.5}$, and Hf$_{0.33}$Zr$_{0.33}$Ti$_{0.33}$, respectively.
    }
    \label{fig:formation_energy}
\end{figure}

We apply Pareto-optimal SQSs to estimate the formation energies of fcc-based and hcp-based perfectly random structures with DFT calculations to show the convergence behavior of the formation energies.
We select fcc-based SQSs for a Cu-Au-Pd system and hcp-based SQSs for a Hf-Zr-Ti system.
In the Cu-Au-Pd system, Cu, Au, and Pd are in the ground state with the fcc structure, and Cu$_{0.5}$Pd$_{0.5}$, Cu$_{0.5}$Au$_{0.5}$, and Au$_{0.5}$Pd$_{0.5}$ are fcc-based solid solutions at temperatures of around several hundred degrees, although B2 and $\mbox{L1}_{0}$ ordered structures have been reported to exist in the Cu-Pd and Cu-Au systems at low temperatures, respectively \cite{villars2016asm}.
Similarly, in the Hf-Zr-Ti system, Hf, Zr, and Ti are in the ground state with the hcp structure at low temperatures, and Hf$_{0.5}$Ti$_{0.5}$, Hf$_{0.5}$Zr$_{0.5}$, and Zr$_{0.5}$Ti$_{0.5}$ are hcp-based solid solutions at low temperatures.

For each system, we calculate the formation energies of equiatomic SQSs in the binary and ternary systems.
DFT calculations are performed using plane-wave basis sets and the projector augmented wave (PAW) \cite{PhysRevB.50.17953,PhysRevB.59.1758} with the Perdew--Burke--Ernzerhof (PBE) exchange-correlation functional \cite{PhysRevLett.77.3865} implemented in the Vienna \textit{ab initio} Simulation Package (\textsc{vasp}) \cite{PhysRevB.47.558,PhysRevB.54.11169,KRESSE199615} (version~5.4.4).
The plane-wave-energy cutoff is set at 400 eV.
For each initial structure, the atomic positions and lattice constants are relaxed until residual forces are less than $10^{-2} \, \mbox{eV} / \mbox{\AA}$.
Electronic structure optimization is performed with smearing with $\sigma = 0.2\, \mathrm{eV}$ in the Methfessel--Paxton scheme \cite{PhysRevB.40.3616} until the energetic change is less than $10^{-4}\, \mathrm{eV} / \mbox{supercell}$.
After structural relaxation, the total energy is calculated by the tetrahedron method with Bl\"{o}chl corrections \cite{PhysRevB.49.16223}.

Figure~\ref{fig:formation_energy} shows the formation energies of SQSs computed by the DFT calculations.
In the upper figure, the formation energies of the fcc-based SQSs in the binary and ternary systems converge around 1 meV/atom with increasing number of sites in SQSs.
Also, the formation energies of the hcp-based SQSs converge around 2 meV/atom in the binary system and around 10 meV/atom in the ternary system.
In the currently selected systems, the convergence behavior can be recognized by including SQSs with a large number of sites such as SQS-27 and SQS-32, although the formation energy converges at a smaller number of sites.
These results indicate that the Pareto-optimal SQSs obtained by the current ZDD-based method closely mimic the perfectly random structure.

\section{Conclusion}
We have presented an efficient ZDD-based algorithm for searching for SQSs, which works for arbitrary lattices.
In the current algorithm, ZDDs are sequentially constructed by imposing constraints for extracting SQSs one by one, and the obtained final ZDD represents a set of only SQSs.
We have also applied the current ZDD-based algorithm to search for fcc-based and hcp-based SQSs in binary, ternary, and quaternary systems.
The current approach extracts only a small number of SQSs from a ZDD composed of many candidate derivative structures (more than $10^{12}$).
Consequently, we propose SQSs that are better optimized than those proposed in the literature.
Therefore, the use of ZDDs significantly improves the efficiency of enumerating derivative structures that satisfy the constraints.
Furthermore, the current algorithm and ideas used to introduce constraints are also helpful for the enumeration of feasible structures satisfying some constraints that are not used in this study.

\begin{acknowledgments}
This work was supported by
    a Grant-in-Aid for JSPS Research Fellows (Grant Number 21J10712),
    a Grant-in-Aid for Scientific Research (B) (Grant Number 19H02419),
    and a Grant-in-Aid for Scientific Research on Innovative Areas (Grant Number 19H05787)
from the Japan Society for the Promotion of Science (JSPS).
\end{acknowledgments}

\appendix
\section{\label{sec:app-table}Pair clusters and development of ZDDs}

The nonequivalent pair clusters for fcc and hcp are tabulated in Tables~\ref{tab:fcc_pair_clusters} and \ref{tab:hcp_pair_clusters}, respectively.
The processes of reducing the number of candidates for fcc-based SQSs are shown in Tables~\ref{tab:number_into_fcc_ternary_sqs} and \ref{tab:number_into_fcc_quaternary_sqs} for concentrations $ABC$ and $ABCD$, respectively.
The processes of reducing the number of candidates for hcp-based SQSs are shown in Tables~\ref{tab:number_into_hcp_binary_sqs}, \ref{tab:number_into_hcp_ternary_sqs}, and \ref{tab:number_into_hcp_quaternary_sqs} for concentrations $AB$, $ABC$, and $ABCD$, respectively.

\begin{table*}[t]
    \centering
    \caption{
        Nonequivalent pair clusters for fcc.
        The unit of sizes of clusters is lattice constant $a$ of an fcc conventional unit cell.
        The second column shows the distance of the $n$th NN pairs.
        The third column shows the number of equivalent pairs in the primitive fcc cell.
        The fourth and fifth columns show Cartesian coordinates of representatives of nonequivalent pairs.
    }
    \label{tab:fcc_pair_clusters}
    \begin{ruledtabular}
    \begin{tabular}{ccccc}
        $n$th neighbor & Size & Multiplicity & Cartesian coordinates 1 & Cartesian coordinates 2 \\ \hline
        1 & 0.7071 & 6  & (0, 0, 0) & (-0.5,  0.0,  0.5) \\
        2 & 1.0000 & 3  & (0, 0, 0) & ( 0.0,  1.0,  0.0) \\
        3 & 1.2247 & 12 & (0, 0, 0) & ( 1.0,  0.5,  0.5) \\
        4 & 1.4142 & 6  & (0, 0, 0) & ( 0.0,  1.0,  1.0) \\
        5 & 1.5811 & 12 & (0, 0, 0) & ( 0.0, -0.5,  1.5) \\
        6 & 1.7321 & 4  & (0, 0, 0) & (-1.0,  1.0, -1.0) \\
        7 & 1.8708 & 24 & (0, 0, 0) & (-1.5, -1.0,  0.5) \\
        8 & 2.0000 & 3  & (0, 0, 0) & ( 0.0,  0.0,  2.0) \\
        9 & 2.1213 & 12 & (0, 0, 0) & (-2.0, -0.5,  0.5) \\
        10 & 2.1213 & 6 & (0, 0, 0) & ( 0.0,  1.5,  1.5) \\
    \end{tabular}
    \end{ruledtabular}
\end{table*}

\begin{table*}[t]
    \centering
    \caption{
        Nonequivalent pair clusters for hcp with the ideal $c/a$ ratio.
        The unit of sizes of clusters is lattice constant $a$ of an hcp conventional unit cell.
    }
    \label{tab:hcp_pair_clusters}
    \begin{ruledtabular}
    \begin{tabular}{ccccc}
        $n$th neighbor & Size & Multiplicity & Cartesian coordinates 1 & Cartesian coordinates 2 \\ \hline
        1 & 1.0000 & 6 & (0.0000, 0.0000, 0.0000) & (-0.5000, 0.2887, -0.8165) \\
        2 & 1.0000 & 6 & (0.5000, 0.2887, 0.8165) & (1.5000, 0.2887, 0.8165) \\
        3 & 1.4142 & 6 & (0.0000, 0.0000, 0.0000) & (1.0000, -0.5774, -0.8165) \\
        4 & 1.6330 & 2 & (0.5000, 0.2887, 0.8165) & (0.5000, 0.2887, 2.4495) \\
        5 & 1.7321 & 6 & (0.5000, 0.2887, 0.8165) & (2.0000, -0.5774, 0.8165) \\
        6 & 1.7321 & 12 & (0.0000, 0.0000, 0.0000) & (-1.0000, 1.1547, -0.8165) \\
        7 & 1.9149 & 12 & (0.0000, 0.0000, 0.0000) & (1.0000, 0.0000, 1.6330) \\
        8 & 2.0000 & 6 & (0.5000, 0.2887, 0.8165) & (1.5000, -1.4434, 0.8165) \\
        9 & 2.2361 & 12 & (0.0000, 0.0000, 0.0000) & (2.0000, -0.5774, 0.8165) \\
        10 & 2.3805 & 12 & (0.0000, 0.0000, 0.0000) & (1.5000, -0.8660, 1.6330) \\
        11 & 2.4495 & 6 & (0.0000, 0.0000, 0.0000) & (2.0000, 1.1547, 0.8165) \\
        12 & 2.5166 & 6 & (0.0000, 0.0000, 0.0000) & (0.5000, 0.2887, 2.4495) \\
        13 & 2.5820 & 12 & (0.0000, 0.0000, 0.0000) & (1.0000, -1.7321, 1.6330) \\
        14 & 2.6458 & 12 & (0.0000, 0.0000, 0.0000) & (-1.5000, 2.0207, -0.8165) \\
        15 & 2.6458 & 12 & (0.0000, 0.0000, 0.0000) & (2.0000, 1.7321, 0.0000) \\
        16 & 2.7080 & 6 & (0.0000, 0.0000, 0.0000) & (0.0000, 1.1547, 2.4495) \\
        17 & 2.8868 & 12 & (0.0000, 0.0000, 0.0000) & (1.0000, 1.1547, -2.4495) \\
        18 & 3.0000 & 6 & (0.5000, 0.2887, 0.8165) & (2.0000, -2.3094, 0.8165) \\
    \end{tabular}
    \end{ruledtabular}
\end{table*}

\begin{table}[t!]
    \centering
    \caption{
        Numbers of labelings in searching for ternary fcc-based SQSs of $ABC$ supercell with 27 sites with transformation matrix $\mathbf{M} = [[1, 0, 0], [0, 3, 0], [1, 3, 9]]$.
    }
    \label{tab:number_into_fcc_ternary_sqs}
    \setlength\tabcolsep{6pt}  %
    \begin{ruledtabular}
    \begin{tabular}{lr}
        Constraints & Number of labelings \\ \hline
        No constraint & 7,625,597,484,987 \\
        + Concentration $ABC$ & 227,873,431,500 \\
        + Nonequivalent & 4,219,878,612 \\
        + Up to the first NN & 7,077,792 \\
        + Up to the second NN & 62,797 \\
        + Up to the third NN & 174 \\
        + Up to the fourth NN & 38 \\
        + Up to the fifth NN & 26 \\
        + Up to the sixth NN & 2 \\
        + Up to the seventh NN & 2 \\
        + Up to the eighth NN & 2 \\
        + Up to the ninth NN & 2 \\
    \end{tabular}
    \end{ruledtabular}
\end{table}

\begin{table}[t!]
    \centering
    \caption{
        Numbers of labelings in searching for quaternary fcc-based SQSs of $ABCD$ supercell with 16 sites with transformation matrix $\mathbf{M} = [[1, 0, 0], [0, 2, 0], [1, 4, 8]]$.
    }
    \label{tab:number_into_fcc_quaternary_sqs}
    \setlength\tabcolsep{6pt}  %
    \begin{ruledtabular}
    \begin{tabular}{lr}
        Constraints & Number of labelings \\ \hline
        No constraint & 4,294,967,296 \\
        + Concentration $ABCD$ & 63,063,000 \\
        + Nonequivalent & 990,906 \\
        + Up to the first NN & 498 \\
        + Up to the second NN & 48 \\
    \end{tabular}
    \end{ruledtabular}
\end{table}

\begin{table}[t!]
    \centering
    \caption{
        Numbers of labelings in searching for binary hcp-based SQSs of $AB$ supercell with 32 sites with transformation matrix $\mathbf{M} = [[1, 0, 0], [1, 4, 0], [1, 1, 4]]$.
    }
    \label{tab:number_into_hcp_binary_sqs}
    \setlength\tabcolsep{6pt}  %
    \begin{ruledtabular}
    \begin{tabular}{lr}
        Constraints & Number of labelings \\ \hline
        No constraint & 4,294,967,296 \\
        + Concentration $AB$ & 165,636,900 \\
        + Nonequivalent & 5,182,744 \\
        + Up to the first NN & 1,018,101 \\
        + Up to the second NN & 104,616 \\
        + Up to the third NN & 19,531 \\
        + Up to the fourth NN & 3,218 \\
        + Up to the fifth NN & 716 \\
        + Up to the sixth NN & 110 \\
        + Up to the seventh NN & 2 \\
        + Up to the eighth NN & 2 \\
        + Up to the ninth NN & 2 \\
        + Up to the tenth NN & 2 \\
        + Up to the eleventh NN & 2 \\
        + Up to the twelfth NN & 2 \\
        + Up to the thirteenth NN & 2 \\
        + Up to the fourteenth NN & 2 \\
        + Up to the fifteenth NN & 2 \\
        + Up to the sixteenth NN & 2 \\
        + Up to the seventeenth NN & 2 \\
        + Up to the eighteenth NN & 2 \\
    \end{tabular}
    \end{ruledtabular}
\end{table}

\begin{table}[t!]
    \centering
    \caption{
        Numbers of labelings in searching for ternary hcp-based SQSs of $ABC$ supercell with 18 sites with transformation matrix $\mathbf{M} = [[1, 0, 0], [0, 1, 0], [0, 4, 9]]$.
    }
    \label{tab:number_into_hcp_ternary_sqs}
    \setlength\tabcolsep{6pt}  %
    \begin{ruledtabular}
    \begin{tabular}{lr}
        Constraints & Number of labelings \\ \hline
        No constraint & 387,420,489 \\
        + Concentration $ABC$ & 2,822,400 \\
        + Nonequivalent & 157,644 \\
        + Up to the first NN & 1,876 \\
        + Up to the second NN & 23 \\
        + Up to the third NN & 12 \\
        + Up to the fourth NN & 12 \\
    \end{tabular}
    \end{ruledtabular}
\end{table}

\begin{table}[t!]
    \centering
    \caption{
        Numbers of labelings in searching for quaternary hcp-based SQSs of $ABCD$ supercell with 16 sites with transformation matrix $\mathbf{M} = [[1, 0, 0], [0, 2, 0], [2, 1, 4]]$.
    }
    \label{tab:number_into_hcp_quaternary_sqs}
    \setlength\tabcolsep{6pt}  %
    \begin{ruledtabular}
    \begin{tabular}{lr}
        Constraints & Number of labelings \\ \hline
        No constraint & 4,294,967,296 \\
        + Concentration $ABCD$ & 6,350,400 \\
        + Nonequivalent & 199,740 \\
        + Up to the first NN & 4,020 \\
        + Up to the second NN & 96 \\
        + Up to the third NN & 96 \\
    \end{tabular}
    \end{ruledtabular}
\end{table}

\section{\label{sec:app-mcsqs}Comparison with stochastic method}

\begin{table}[tb]
    \centering
    \caption{
        Formation energies of binary fcc-based SQSs with 40 sites of Cu$_{0.5}$Au$_{0.5}$ (units: meV/atom) .
        The first column corresponds to the formation energy of SQS-40 generated by the ZDD-based method shown in Table~\ref{tab:result_1_sqs_fcc}.
        The other columns correspond to the formation energies of SQSs generated by the stochastic method with different random seeds, respectively.
    }
    \label{tab:result_4_fcc_CuAu}
    \begin{ruledtabular}
        \begin{tabular}{ccccc}
            ZDD & Stochastic 1 & Stochastic 2 & Stochastic 3 & Stochastic 4 \\ \hline
            -15.60 & -15.54 & -15.97 & -13.30 & -16.08
        \end{tabular}
    \end{ruledtabular}
\end{table}

\begin{table}[tb]
    \centering
    \caption{
        Formation energies of ternary fcc-based SQSs with 27 sites of Cu$_{0.33}$Pd$_{0.33}$Au$_{0.33}$ (units: meV/atom) .
    }
    \label{tab:result_4_fcc_CuPdAu}
    \begin{ruledtabular}
        \begin{tabular}{ccccc}
            ZDD & Stochastic 1 & Stochastic 2 & Stochastic 3 & Stochastic 4 \\ \hline
            -86.66 & -87.36 & -85.04 & -81.18 & -80.93
        \end{tabular}
    \end{ruledtabular}
\end{table}

Here, we compare SQSs generated by the ZDD-based method and the stochastic method.
We generate fcc-based SQSs with 40 sites for composition $AB$ and fcc-based SQSs with 27 sites for composition $ABC$ using a stochastic method implemented in \textsc{atat} (\textsc{mcsqs} package \cite{mcsqs}).
They are comparable with SQSs listed in Table.~\ref{tab:result_1_sqs_fcc}.
The stochastic method aims to solve the following minimization problem,
\begin{align}
    \label{eq:mcsqs-problem}
    \min_{ \mathbf{c} \in \mathcal{C}_{\mathrm{conc}} }
    \left(
        \sum_{\alpha} \left| \Pi_{\alpha}(\mathbf{c}) - \overline{\Pi}_{\alpha} \right| -\omega L(\mathbf{c})
    \right),
\end{align}
where the second term is the penalty function, $L(\mathbf{c})$ is the size of the largest cluster among perfectly matched clusters with the perfectly random structure, and $\omega$ is a positive regularization coefficient.
The penalty function is introduced so that the correlation functions of small clusters become exactly the same as those of the perfectly random structure.
Note that the current ZDD-based method does not introduce such a regularization term explicitly.
Instead, the current method uses constraints that the correlation functions of pair clusters must be exactly the same as those of the perfectly random structure.

In the function of Eq.~\eqref{eq:mcsqs-problem}, we set the maximum pair cluster to the tenth NN and ninth NN pairs for obtaining SQSs of compositions $AB$ and $ABC$, respectively.
To minimize the function of Eq.~\eqref{eq:mcsqs-problem}, we perform simulated annealing implemented in \textsc{atat} within approximately $10^{5}$ seconds using four different random seeds for each composition.  %
For composition $AB$, the stochastic method finds the same structures as the perfectly random structure in terms of all pairs up to the sixth NN.
For composition $ABC$, it finds the same structures as the perfectly random structure in terms of all pairs up to the second NN.
Thus, the stochastic method generates SQSs that are not well optimized even when the ZDD-based method can find the well-optimized SQS from the enormous search space.

Then, we compare the formation energies of the SQSs generated by the stochastic method and the ZDD-based method in Cu$_{0.5}$Au$_{0.5}$ and Cu$_{0.33}$Pd$_{0.33}$Au$_{0.33}$.
DFT calculations are performed with the same computational procedure shown in Sec.~\ref{sec:results-realistic-systems}.
Table~\ref{tab:result_4_fcc_CuAu} lists the formation energies of SQSs with 40 sites for Cu$_{0.5}$Au$_{0.5}$.
As shown in Table~\ref{tab:result_4_fcc_CuAu}, the formation energy depends on the random seed used in the stochastic method.
The stochastic method shows a variation of approximately 3 meV/atom in the formation energy.
Table~\ref{tab:result_4_fcc_CuPdAu} lists the formation energies of SQSs with 27 sites for Cu$_{0.33}$Pd$_{0.33}$Au$_{0.33}$.
The stochastic method involves a variation of approximately 6 meV/atom in the formation energy.
The variations of the formation energy in the stochastic method are significant compared with the converged value of the formation energy obtained by the ZDD-based method.
Although the variations in the stochastic method and convergence behavior of the formation energy in the ZDD-based method depend on the alloy system, the current results indicate that the present ZDD-based method is useful for generating high-quality SQSs.

\bibliography{references}%

\end{document}